\documentclass[fleqn,usenatbib]{mnras}

\usepackage{newtxtext,newtxmath}
\usepackage{xcolor}

\usepackage[T1]{fontenc}

\DeclareRobustCommand{\VAN}[3]{#2}
\let\VANthebibliography\thebibliography
\def\thebibliography{\DeclareRobustCommand{\VAN}[3]{##3}\VANthebibliography}

\newcommand{\comment}[1]{}

\usepackage{graphicx}	
\usepackage{amsmath}	
\usepackage{multirow}

\title[Mimicking the halo--galaxy connection using machine learning]{Mimicking the halo--galaxy connection using machine learning}

\author[de Santi, N. S. M. et al.]{
\parbox[t]{\textwidth}{
Natalí S. M. de Santi$^{1}$,\thanks{E-mail: natalidesanti@gmail.com}
Natália V. N. Rodrigues$^{1}$,\thanks{E-mail: natvnr@gmail.com}
Antonio D. Montero-Dorta$^{2}$,
L. Raul Abramo$^{1}$, 
Beatriz Tucci$^{1,3}$, 
M. Celeste Artale$^{4,5,6}$}
\\
\\
$^{1}$Instituto de Física, Universidade de São Paulo, 
R. do Matão 1371, 05508-090, São Paulo, SP, Brazil\\
$^{2}$Departamento de Física, Universidad Técnica Federico Santa María, Casilla 110-V, Avda. España 1680, Valparaíso, Chile.\\
$^{3}$Max–Planck–Institut für Astrophysik, Karl–Schwarzschild–Straße 1, 85748 Garching, Germany\\
$^{4}$ Institut für Astro- und Teilchenphysik, Universität Innsbruck, Technikerstrasse 25/8, 6020 Innsbruck, Austria\\ 
$^{5}${Department of Physics and Astronomy, Purdue University, 525 Northwestern Avenue, West Lafayette, IN 47907, USA}\\
$^{6}$ Physics and Astronomy Department Galileo Galilei, University of Padova, Vicolo dell’Osservatorio 3, I-35122, Padova, Italy
}

\date{Accepted XXX. Received YYY; in original form ZZZ}

\pubyear{2021}

\begin{document}
\label{firstpage}
\pagerange{\pageref{firstpage}--\pageref{lastpage}}

\maketitle

\begin{abstract}

Elucidating the connection between the properties of galaxies and the properties of their hosting haloes is a key element in galaxy formation. When the spatial distribution of objects is also taken under consideration, it becomes very relevant for cosmological measurements. In this paper, we use machine learning techniques to analyse these intricate relations in the IllustrisTNG300 magnetohydrodynamical simulation,  predicting baryonic properties from halo properties.
We employ four different algorithms: {\it{extremely randomized trees}}, {\it{K-nearest neighbours}}, {\it{light gradient boosting machine}}, and {\it{neural networks}}, along with a unique and powerful combination of the results from all four approaches.
Overall, the different algorithms produce consistent results in terms of predicting galaxy properties from a set of input halo properties that include halo mass, concentration, spin, and halo overdensity. For stellar mass, the 
Pearson correlation coefficient is 0.98, dropping down to 0.7-0.8 for specific star formation rate (sSFR), colour, and size. 
In addition, we apply, for the first time in this context, an existing data augmentation method, 
{\it{synthetic minority over-sampling technique for regression with Gaussian noise}} (SMOGN),
designed to alleviate the problem of imbalanced data sets, 
showing that it improves the overall shape of the predicted distributions
and the scatter in the halo--galaxy relations.
We also demonstrate that our predictions are good enough to reproduce the power spectra of multiple galaxy populations, defined in terms of stellar mass, sSFR, colour, and size with high accuracy. Our results align with previous reports suggesting that certain galaxy properties cannot be reproduced using halo features alone.  

\end{abstract}




\section{Introduction}

Illuminating the intricate relations between the baryonic and dark matter components of the Universe has progressively become one of the most important areas of research in cosmology and galaxy evolution. The question is currently centred on characterising the connections between the properties of galaxies and the properties of the dark-matter haloes where they form and evolve, in the cosmological context of the large-scale structure (LSS) of the Universe. 

Several methods are currently employed in order to investigate and characterise the aforementioned connection (see a comprehensive review on halo--galaxy connection methods in \citealt{Wechsler2018}). Empirical techniques include {\it{sub-halo abundance matching}} (SHAM, e.g., \citealt{Conroy2006,Behroozi2010,Trujillo-Gomez2011,Favole2016,Guo2016,Contreras2020,Contreras2021,Hadzhiyska2021, Favole2021}), {\it{halo cccupation distributions}} (HODs, e.g., \citealt{Berlind2002, Zehavi2005,Zehavi2018,Artale2018,Bose2019,Hadzhiyska2020B,Xu2021}) and {\it{empirical forward modeling}} (e.g., \citealt{becker2015_eam, moster2018_emerge, behroozi2019_um}). Physical models such as {\it{semi-analytic models}} (SAMs, e.g., \citealt{White1991, Guo2013}) and {\it{hydrodynamical simulations}} (e.g., \citealt{Somerville2015,Naab2017,Pillepich2018b,Pillepich2018, Springel2018}) are computationally more challenging, but they aim for a more detailed understanding of the physical processes that shape the halo--galaxy link. Among these methods, the latter, which employ known physics to simulate, at a sub-grid level, a variety of processes that are related to galaxy formation (e.g., star formation, radiative metal cooling, and supernova, stellar, and black hole feedback) are by far the most sophisticated modelling tools available to the community (see reviews in \citealt{Somerville2015,Naab2017}).

Despite the abundance of techniques, it has become progressively more evident in recent years that dissecting the complexity of the halo--galaxy connection requires the development of new approaches, even at the data handling level. In this context, machine learning (ML) techniques are emerging as a promising avenue, as they are proven to be a powerful tool for unveiling hidden relations between variables in very complex systems, in a variety of scientific fields \footnote{Note that, in this context, ML techniques could be considered a halo--galaxy connection technique by itself, as the result of the analysis is a set of models that characterise this link.}. Among the applications of ML in astrophysics, we can cite the search for exoplanets \citep{Dattilo2019}, the generation of weak lensing convergence maps \citep{Mustafa2019}, the estimation of photometric redshifts for current and future galaxy surveys (e.g., \citealt{D_Isanto_2018}), or the construction of cosmological hydrodynamical simulations (e.g., \citealt{VillaescusaNavarro2021}), to name but a few.  

In this work, we test the performance of several ML methods by applying them to the problem of predicting the central galaxy -- halo connection. 
For this task, the IllustrisTNG  magnetohydrodynamical cosmological simulation \citep{Pillepich2018b,Pillepich2018} appears, as it has been extensively demonstrated previously, as the perfect test bench. The link between each galaxy and its dark matter halo is known a priori, but the model is complex enough to produce a highly realistic data set.  ML techniques have been employed in the context of the halo--galaxy connection before 
(e.g., \citealt{Kamdar2016, Agarwal2018, Calderon2019, Jo2019, Man2019,Yip2019, Zhang2019,Jo2019,Kasmanoff2020, Delgado2021,McGibbon2021,Lovell2021,shao2021}). What has generally transpired from these analyses is that stellar mass is by far the galaxy property that is easier to predict, as it is known to display a strong connection with halo mass. Other properties, such as colour or star formation rate (SFR), which are severely influenced by several secondary halo properties in non-trivial ways, are still challenging to reproduce even by the most sophisticated ML methods. To give some numbers, the Pearson correlation coefficient between the true and the predicted values for stellar mass are often above 0.9 in these analyses (e.g., \citealt{Kamdar2016,Lovell2021}), whereas the same metric yields values typically below 0.8 for other properties such as SFR (e.g., \citealt{Kamdar2016, Agarwal2018,Lovell2021}). 

In the context of the Illustris simulations, in particular, \cite{Kamdar2016} used the {\it{extremely randomized trees}} (ERT) ML method to predict several (simulated) galaxy properties such as gas mass, stellar mass, black hole mass, SFR, or colour. More recently, several other works have focused their efforts on predicting stellar mass alone, using {\it{convolutional neural networks}} applied to the improved IllustrisTNG simulation (see, e.g., \citealt{Zhang2019, Yip2019, Kasmanoff2020}). These authors have extended the scope of the analyses to the measurement of clustering properties and the comparison with halo--galaxy connection techniques such as HODs. Also noteworthy is the study presented in \cite{Jo2019}, where the ERT technique is applied to IllustrisTNG in order to model the connection between halo properties and  
both stellar mass and SFR. The trained machine is then applied to the MultiDark dark-matter-only simulation and results are compared with a SAM. 

In this paper, we use the largest IllustrisTNG simulation box (hereafter, TNG300, of 205 $h^{-1}$Mpc of side length) to evaluate the performance of the following ML methods in the context of galaxy-property prediction: {\it{extremely randomized trees}} (ERT), {\it{K-nearest neighbours}} (kNN), {\it{light gradient boosting machine}} (LightGBM or LGBM, hereafter), and {\it{neural networks}} (NN). These different ML flavours are tested, following a standard procedure, in terms of predicting the following fundamental properties for central galaxies: stellar mass, half-mass radius, specific star formation rate (sSFR) and colour. On the halo side, the link towards galaxies is established on the basis of five halo properties, namely: halo mass, concentration, spin, age, and the overdensity halo environment proxy. 
Instead of choosing a single method, we combine the predictions from different methods and use them as another data set that feeds a different ML method to provide a new, combined prediction of each galaxy property. These ``stacked'' predictions therefore take into account the positive and negative aspects of each individual model.

An important part of our analysis involves testing the {\it{synthetic minority over-sampling technique for regression with Gaussian noise}} (SMOGN) data  augmentation technique \citep{pmlr-v74-branco17a}. This method was developed as a means of alleviating the classic ML problem of imbalanced data sets, i.e., situations where some underrepresented subsets of the data are considered as important (or interesting for the task in hand) as other highly-populated regions of the parameter space. This situation is currently common in halo--galaxy connection models, which aim at reproducing multiple galaxy populations with high precision (see an alternative approach in  \citealt{Jo2019}). For previous applications of the SMOGN technique, see
\cite{Lu2021} and \cite{Sharif2021}, in the context of satellite observations and continuous integration testing, respectively.

The paper is structured as follows. The IllustrisTNG data, including the halo and galaxy properties employed in the analysis and our data pre-selection scheme, are presented in Section \ref{data}. Section \ref{methodology} describes our methodology, which is based, as mentioned above, in several ML algorithms and techniques. The main results of the paper in terms of data prediction, feature importance, and the power spectrum measurement are presented in Section \ref{results}. Finally, in Section \ref{discussion}, we discuss the implications of our results, explain future plans, and provide a brief summary of our results. The IllustrisTNG simulation adopts the standard $\Lambda$CDM cosmology \citep{Planck2016}, with parameters $\Omega_{\rm m} = 0.3089$,  $\Omega_{\rm b} = 0.0486$, $\Omega_\Lambda = 0.6911$, $H_0 = 100\,h\, {\rm km\, s^{-1}Mpc^{-1}}$ with $h=0.6774$, $\sigma_8 = 0.8159$, and $n_s = 0.9667$.

\begin{figure*}
    \centering
    \includegraphics[scale=0.65]{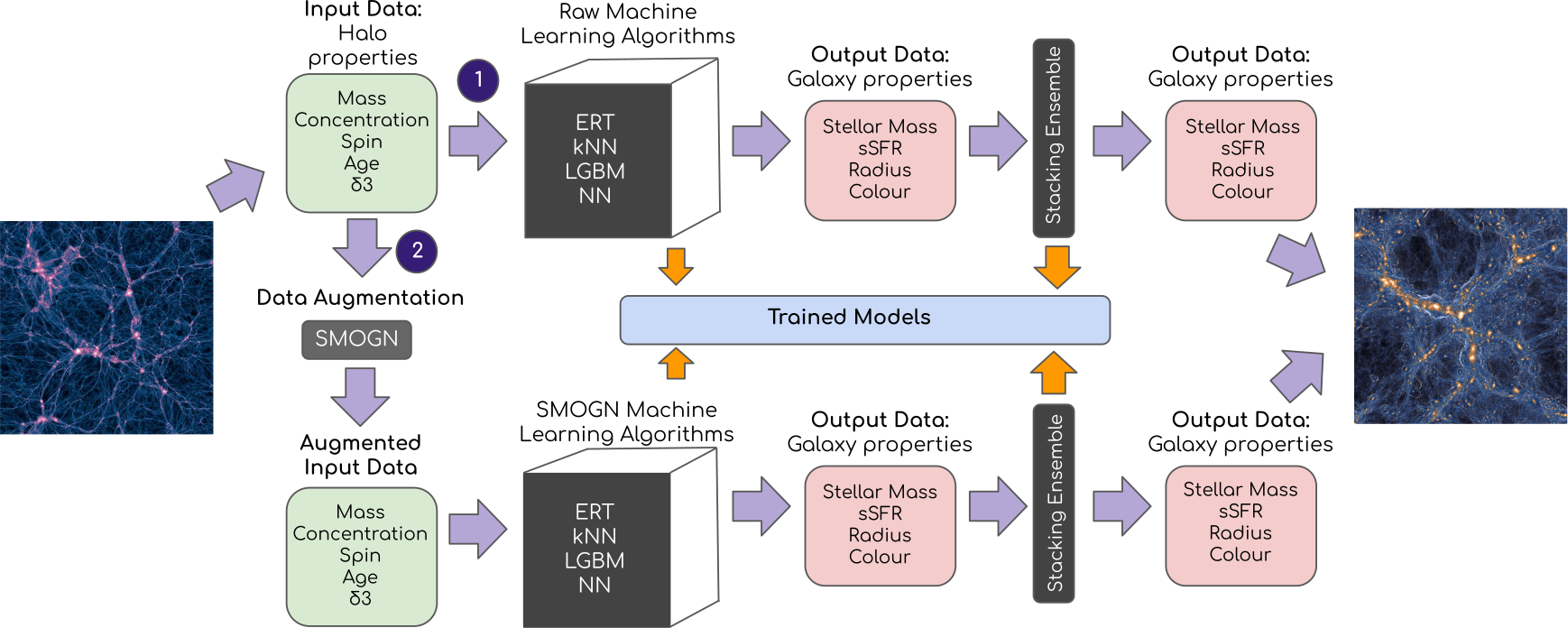}
    \caption{A schematic summary of the methodology followed in this analysis. First, a data pre-selection is performed in order to generate the input data set. Once this initial catalogue is constructed, our method takes two different paths: in Path 1 (top), several ML algorithms are applied (ERT, kNN, LGBM, and NN) to the input data set; these are the ``raw'' models. A separate approach, which employs the SMOGN data augmentation technique, is shown in Path 2 (bottom). Subsequently, the same ML methods are applied to the SMOGN input data set. Both paths result in trained models for the galaxy properties. Finally, we have also implemented a stacking ML technique separately for each path, where all ML methods for the corresponding path are combined. The final output data set comprises our predictions for the galaxy properties under analysis.}
    \label{fig:general_picture}
\end{figure*}

\section{The Illustris-TNG data}
\label{data}

Our analysis is based on data from the IllustrisTNG magnetohydrodynamical cosmological simulation \citep{Pillepich2018b,Pillepich2018,Nelson2018_ColorBim,Nelson2019,Marinacci2018,Naiman2018,Springel2018}. The IllustrisTNG simulation suite was produced using the {\sc arepo} moving-mesh code \citep{Springel2010} and is considered an improved version of the previous Illustris simulation \citep{Vogelsberger2014a, Vogelsberger2014b, Genel2014}. The updated IllustrisTNG sub-grid models account for star formation, radiative metal cooling, chemical enrichment from SNII, SNIa, and AGB stars, stellar feedback, and super-massive black hole feedback. These models were calibrated to reproduce a set of observational constraints that include the observed $z=0$ galaxy stellar mass function, the cosmic SFR density, the halo gas fraction, the galaxy stellar size distributions, and the black hole -- galaxy mass relation.

In this work, we model the halo--galaxy using ML techniques. We subsequently measure large-scale halo/galaxy clustering in order to test the accuracy of our modelling. For this reason, 
we chose to analyse the largest box available in the database, IllustrisTNG300-1 (hereafter, TNG300). TNG300 spans a side length of $205\,\,h^{-1}$Mpc and includes periodic boundary conditions. The TNG300 run followed the dynamical evolution of 2500$^3$ dark matter (DM) particles of mass $4.0 \times 10^7$ $h^{-1} {\rm M_{\odot}}$ and (initially) 2500$^3$ gas cells of mass $7.6 \times 10^6$ $h^{-1} {\rm M_{\odot}}$. This box is a useful tool for galaxy formation and clustering science that has proven capable of reproducing a number of observational measurements
(see, e.g., \citealt{Springel2018,Pillepich2018,Bose2019,Beltz-Mohrmann2020,Contreras2020,Gu2020,Hadzhiyska2020,MonteroDorta2020,Shi2020,Hadzhiyska2021,MonteroDorta2021A,MonteroDorta2021B,Favole2021}).

DM haloes in IllustrisTNG are identified using a friends-of-friends (FOF) algorithm with a linking length of 0.2 times the mean inter-particle separation \citep{Davis1985}. The gravitationally bound substructures that we call subhaloes are in turn identified using the {\sc subfind} algorithm \citep{Springel2001,Dolag2009}. Subhaloes containing a non-zero stellar mass component are labelled galaxies.

\subsection{Halo and Galaxy Properties}

In this work, we use both halo and galaxy properties from the TNG300 simulation box. For haloes, the following properties are considered:

\begin{itemize}
  \item Virial mass, $M_{\rm vir}$ [$h^{-1} {\rm M_{\odot}}$], computed by adding up the mass of all gas cells and particles contained within a sphere of radius $R_{\rm vir}$. This sphere is defined so that the enclosed density equals 200 times the critical density. 

  \item Age, described in terms of a formation redshift, $z_\text{1/2}$, defined as the redshift at which half of the present-day halo mass has been accreted into a single subhalo for the first time. For this computation, we use the progenitors of the main branch of the subhalo merger tree computed with {\sc sublink}, which is initialized at $z = 6$. 
  
  \item Spin, $\lambda_{\rm halo}$, defined as in \cite{Bullock2001_2}:

  \begin{equation}\label{eq:spin}
    \lambda_{\rm halo} = \frac{ |J|}{\sqrt{2} { M_{\rm vir}} {V_{\rm vir}} { R_{\rm vir}}},
  \end{equation}
  where J is the angular momentum of the halo and $V_{\rm vir}$ is its circular velocity at the 
  virial radius $R_{\rm vir}$. 

  \item Concentration, $c_{\rm vir}$, defined in the standard way as:
  
  \begin{equation}
   c_{\rm vir} = \frac{R_{\rm vir}}{R_{\rm s}},
  \end{equation}
	where $R_{\rm s}$ is the scale radius, derived from fitting the dark matter density profiles of individual haloes with a NFW profile \citep{nfw1997}.
	
  \item Overdensity on a 3 $h^{-1}$Mpc scale, $\delta_3$, defined as the number density of subhaloes within a sphere of radius $R = 3 h^{-1}$Mpc, normalized by the total number density of subhaloes in the TNG300 box \citep[e.g.,][]{Artale2018,Bose2019}. 
\end{itemize}

Galaxies (i.e., subhaloes with non-zero stellar components in TNG300), are in turn characterized using the following basic properties:

\begin{itemize}
    \item Star formation rate, SFR [${\rm yr^{-1} M_{\odot}}$], computed as the sum of the star formation rate of all gas cells contained in each subhalo.
    \item Galaxy (g-i) colour, derived from the magnitudes provided at the IllustrisTNG database. These magnitudes are computed by adding up the luminosities of all stellar particles of each subhalo (see \citealt{Buser1978}). The IllustrisTNG magnitudes are intrinsic, i.e., the attenuation produced by dust is not included.
    \item Stellar mass, $M_\ast{}$ [$h^{-1} {\rm M_{\odot}}$], defined here as the total mass of all stellar particles bound to each subhalo.
    \item Specific star formation rate, sSFR [$ {\rm yr^{-1}} h$], defined simply as the SFR per unit stellar mass: sSFR = SFR/$M_\ast{}$.
    \item Stellar (3D) half-mass radius, $R_{1/2}^{(*)}$ [$h^{-1} \, {\rm kpc}$], defined as the comoving radius containing half of the stellar mass of each subhalo.
    
\end{itemize}

The above properties are chosen so that they provide a fair description of the structure and assembly history of galaxies and haloes. In the following section we will describe the selection cuts adopted for these properties. 

\subsection{Data pre-selection}
\label{data_pre_selection}

As mentioned previously, only central galaxies are considered in our analysis, which significantly simplifies the modelling of the halo--galaxy connection from TNG300. In addition, and in order to avoid the risk of biasing our results by including unphysical values of the properties presented in the previous section, we have imposed several cuts in the data. First, only haloes with masses above $\log_{10}(M_{\rm vir}$ [$h^{-1} {\rm M_{\odot}}]) = 10.5$ are considered.  Second, a minimum of $\log_{10}(M_{\ast{}}$ [$h^{-1} {\rm M_{\odot}}]) = 8.75$ is imposed for the stellar mass. 
These cuts ensure that haloes have more than 500 dark matter particles and galaxies contain at least 50 stellar mass particles. The final galaxy sample contains $174, 527$ objects.

Both the SFR and the sSFR present a challenge for our analyses in TNG300, since a fraction of 14$\%$ of the total number of galaxies at $z=0$ display a value for  SFR strictly equal to 0. Note that this condition does not represent the demarcation for quiescent galaxies, which is typically set at $\log_{10} ({\rm sSFR}[ {\rm yr}^{-1} h]) \sim -10.5$ in TNG300. To try to prevent any trivial numerical issues caused by this subset of null-SFR galaxies, we have assigned them an artificial SFR by randomly taking a value from a Gaussian distribution of the form $\mathcal{N} (\mu = - 13.5, \sigma = 0.5)$ (see a similar approach in \citealt{Favole2021}). Although this prescription ensures a well defined sSFR for all galaxies, it is still difficult, as discussed in the following sections, to statistically predict the assigned values. Further investigation will be devoted to improve this part of the analysis. 

\begin{figure*}
    \centering
    \includegraphics[scale=0.45]{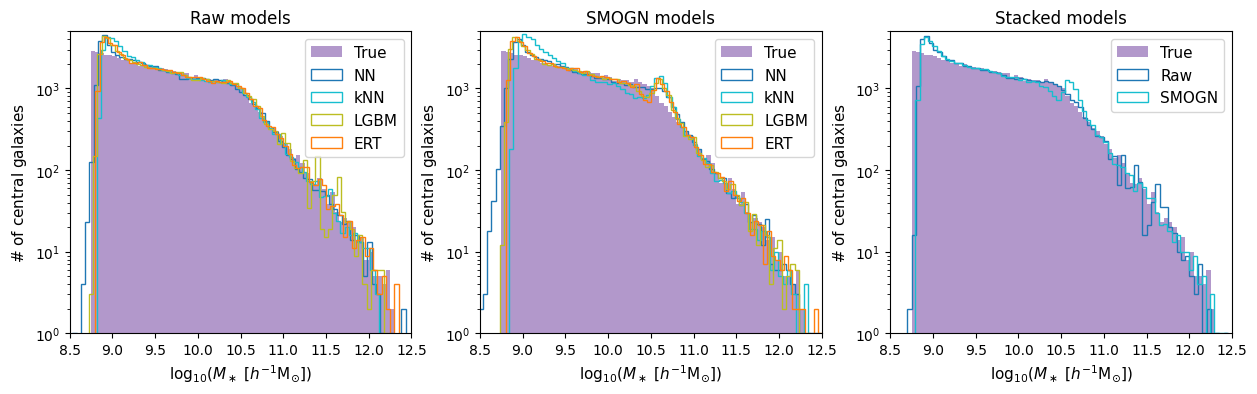}
    \includegraphics[scale=0.45]{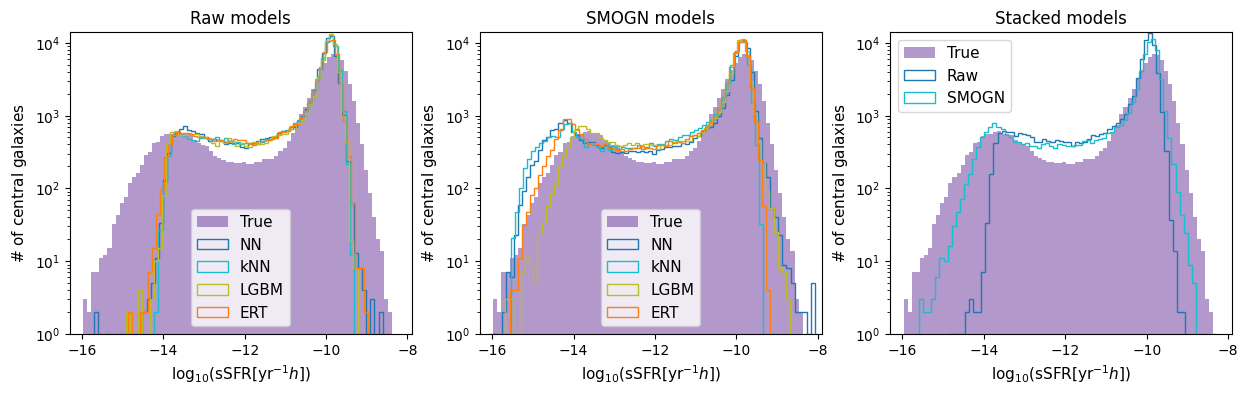}
    \includegraphics[scale=0.45]{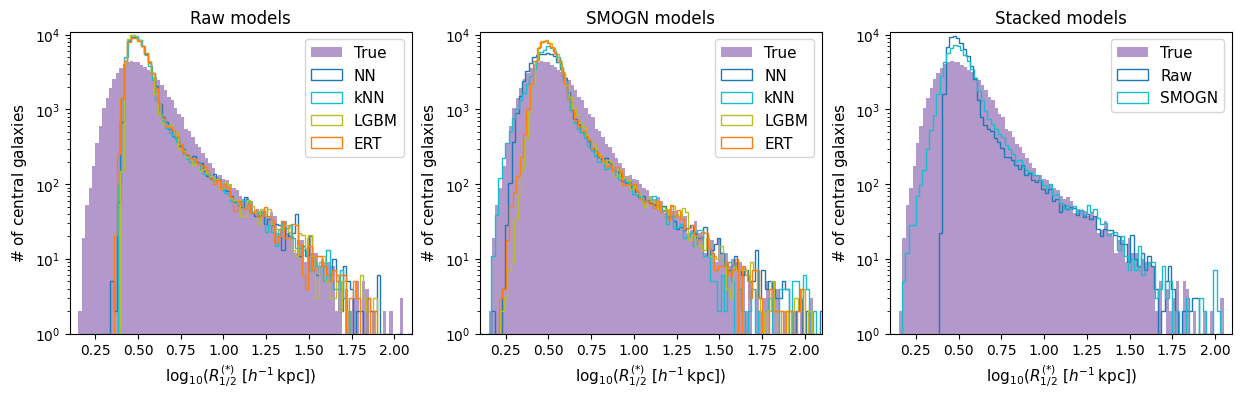}
    \includegraphics[scale=0.45]{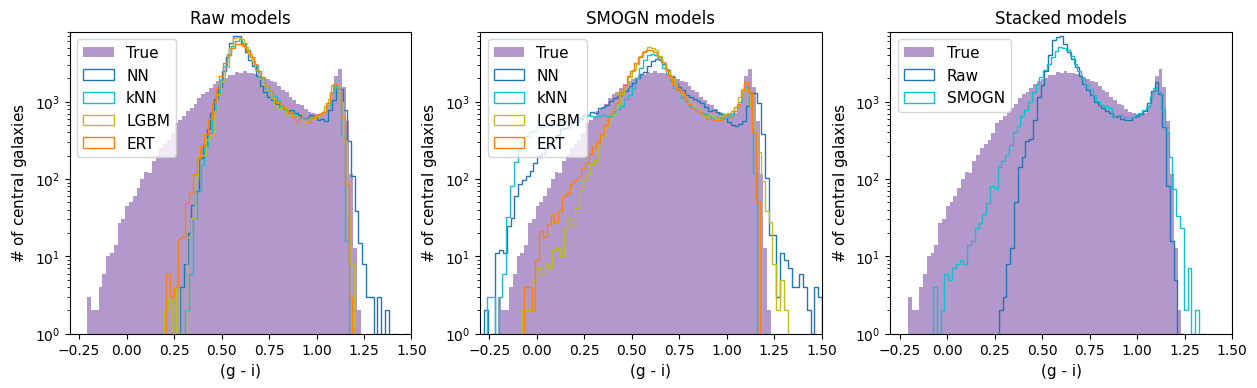}
    \caption{A comparison of the histograms for the true and predicted galaxy properties for all ML models. Each row corresponds to a single galaxy property (stellar mass, sSFR, radius, and colour). The first column corresponds to the ``raw'' ML models (i.e., ML models trained with the ``raw'' TNG300 data), the second column corresponds to the SMOGN models (i.e., the ML models trained with the augmented data, after using SMOGN), and the third column shows the distributions for the stacked models.}
    \label{fig:histograms}
\end{figure*}

\section{Methodology}
\label{methodology}

The goal of this analysis is to reproduce the properties of central galaxies ({\it{target sample}}) in TNG300, using the properties of haloes ({\it{input sample}}). The task of mapping input into output data in terms of continuous variables can be translated into a {\it{machine supervised learning problem}}, more specifically, a regression ML problem\footnote{Supervised learning models are widely used in regression problems, since they are able not only to predict both linear and non-linear relations between continuous variables, but also to recover, to some extent, features such as data distribution and dispersion  \citep{chollet2017, ivezic2014statistics, Awad2015}.}. In essence, the ML algorithm is capable of learning how to best combine the input properties in order to produce realistic output values corresponding to a particular galaxy property. 

Our methodology, which comprises several steps, is presented schematically in Fig. \ref{fig:general_picture}. First, the input data is filtered in the fashion described in Section \ref{data_pre_selection}. 
Once the input data set is constructed, our method takes two different paths: in Path 1 (at the top of Fig. \ref{fig:general_picture}), some of the better established ML algorithms (ERT, kNN, LGBM and NN -- see Section \ref{ML_methods}) are applied to the input data set -- we name these ``raw'' models. A separate approach, which employs the SMOGN data augmentation technique (see Section \ref{sec:SMOGN}), is shown in Path 2 (at the bottom of Fig. \ref{fig:general_picture}). In that approach, the data is first augmented and, subsequently, the same ML methods are applied to the SMOGN input data. Both paths result in trained models for the galaxy properties, and are provided as part of this work. Finally, in addition to the individual models described in Fig. \ref{fig:general_picture}, we have also implemented a stacking ML technique separately for each path, where all ML methods for the corresponding path are combined.

In the remainder of this section, we will briefly describe the ML algorithms employed in the analysis, along with the SMOGN technique and our performance evaluation metrics. 

\begin{figure}
    \centering
    \includegraphics[scale=0.6]{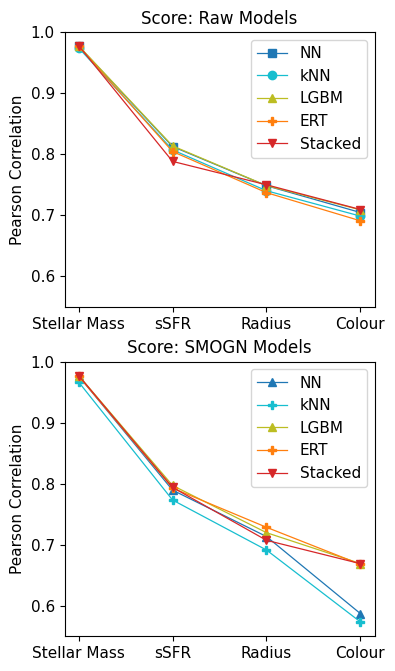}
    \caption{Pearson correlation coefficient for each ML method and galaxy property. The upper plot corresponds to the raw models, while the lower plot displays the SMOGN models. Note that we have opted to connect the dots despite the fact that the properties in the x-axis are not correlated. This format is employed in order to facilitate the readability of the plots.}
    \label{fig:model_score}
\end{figure}

\subsection{Machine Learning Algorithms}
\label{ML_methods}

In order to train the ML models and evaluate its predictions properly, we have split our data
sample in the usual way, i.e., into training, validation and test subsets. 
The training subset is used to train the ML models, i.e., the algorithm uses a slice of data to
learn about the mapping $F(\mathbf{x} \to y)$, and to return a model from which the predictions
can be obtained. 
To avoid overfitting the data, the validation sample is used to monitor how well the model can be generalized to objects outside the training sample, and to help determine the set of 
hyperparameters.
The validation sample is also used to train the stacking ensemble method (see \S\ref{stack model}). 
The test subset remains completely separate from the other two, as it is only used to make the 
final analysis and to evaluate the performance of the methods. 
Our complete data set contains 174,527 objects (also known in the ML nomenclature as 
``instances''). 
From this sample, the fraction of objects corresponding to the training, validation and test 
subsets are $\sim 48\%$, $12\%$, and $40\%$, respectively. Note that having sizeable subsets is 
important for our clustering analysis (specially for the test subset, 
where it is essential to lower the level of shot noise in the power spectrum as much as possible). 

As mentioned before, results using four different techniques are presented in this work, namely: 
ERT, kNN, LGBM, and NN. 
This choice is based on the fact that all these methods have been optimized for their speed in 
the training process, and also because they represent different approaches to solving common 
tasks such as regression. 
In addition, we use, for the first time in the halo--galaxy connection context, both the ML stacking approach and the SMOGN augmentation method.
Below we provide a brief description of each individual technique.

\subsubsection{Extremely randomized trees (ERT)}

ERT is an ensemble method where individual ``weak'' learners (in this case, decision trees, DTs) 
are combined to build a powerful estimator. While a single DT is likely to overfit, ERT 
randomizes the splitting process of the individual DTs, thus reducing the variance of the 
estimator. 
The final prediction of the model is the average over the predictions from all individual DTs -- 
for a more detailed description of the method see \citealt{ert_paper}.
ERTs have already been successfully employed in the context of halo--galaxy connection studies \citep{Kamdar2016, Jo2019, Lovell2021}. 
Here, we use the \href{https://scikit-learn.org/stable/modules/generated/sklearn.ensemble.ExtraTreesRegressor.html}{\texttt{sklearn ensemble Extra Trees Regressor}} library \citep{scikit-learn}.

\subsubsection{K-Nearest Neighbours (kNN)}

kNN is a non-parametric learning algorithm that calculates the distance from a new data point to 
all other training points, assigning the point to the class to which the majority of the $k$ 
neighbours belong. It is predominantly used as an unsupervised ML method, i.e., as a clustering 
algorithm, but can also be framed as a supervised ML method in order to tackle classification and
regression problems.
In the case of regression problems, the final predictions are given by local interpolation of the targets associated with the nearest neighbour -- for more details on the kNN algorithm and its applications see \citet{scikit-learn, ivezic2014statistics}.
As the first presented method, and due to its simplicity, this method has been used to deal with these problems in \cite{Xu2013, Agarwal2018}. Here,  we employ the 
\href{https://scikit-learn.org/stable/modules/generated/sklearn.neighbors.KNeighborsRegressor.html}{\texttt{sklearn K Neighbours Regressor}} library \citep{scikit-learn}.

\subsubsection{Light Gradient Boosting Machine (LGBM)}

LGBM is a gradient boosting framework that implements gradient boosting decision trees (GBDTs). 
GBDTs have been used in a variety of applications, including in astronomy and astrophysics \citep{Lucie2019, golob2021, Li2021, Carvajal2021}. 
Similarly to ERT, the building blocks of GBDTs are DTs, but in this case they are not grown 
independently. Instead, each new DT is an improvement upon the previous one.  
This is achieved by minimising a loss function, which we choose to be the Mean Squared Error (MSE) -- see Eq. \eqref{fig:model_score} (this improvement corresponds to the boosting). 
For further information see \citet{lgbm_paper}.

\subsubsection{Neural Networks (NN)}

NN are a collection of nodes (neurons) that are arranged in a series of layers. Each node of each
layer is connected to all nodes of the next layer through a numeric transformation called
activation function (e.g., ReLu, Sigmoid, and others).
These connections carry weights that, during the training process, are adjusted to minimize the
difference between the network predictions, $y^{\text{pred}}$, and the target values,
$y^{\text{true}}$, quantified in terms of the loss function, which we chose to be the MSE -- see Eq. \eqref{fig:model_score}. For more details on this algorithm, see, e.g., \citealt{chollet2017, bishop1995neural}.
We use here the \href{https://keras.io/}{\texttt{keras}} library \citep{chollet2015}. 
For other works using this method, see \cite{Calderon2019, shao2021}.

\subsubsection{Stacking ensemble model}\label{stack model}

Ensemble ML algorithms employ the outputs of different learning methods in order to improve upon 
the predictions of the individual methods. 
Stacking regression is one of the ensemble methods that can be used to form linear combinations 
of different predictors 
\citep[see, e.g.,][for a detailed description of these algorithms]{Breiman1996}.
In this work, we combine the predictions from the four individual ML methods using the 
\href{https://scikit-learn.org/stable/modules/generated/sklearn.linear_model.LinearRegression.html}{\texttt{LinearRegression}} from \href{https://scikit-learn.org/stable/}{\texttt{scikit-learn}} \citep{scikit-learn}. 

\subsection{Synthetic Minority Over-Sampling Technique for Regression with Gaussian Noise (SMOGN)}
\label{sec:SMOGN}

In ML, there is a common problem with so-called {\it{imbalanced data sets}}. In these samples,
relatively underrepresented regions of the data space may carry the same, or even higher importance, as
it pertains to the scientific context of the analysis, than the bulk of the distribution. These data
are harder to predict by the machine, which focuses on learning about the regions around the peak of
the 
distribution in the parameter space. In regression problems, there are two main ways of dealing 
with the aforementioned issue: pre-processing and model processing techniques. The former ones mainly
focus on applying over-sampling and/or under-sampling techniques, i.e., increasing/reducing the amount
of data in conveniently chosen regions of the distribution.
The scientific importance of these regions can also be weighted, and that is precisely the approach followed by the second methods above. Here we use the SMOGN method, which, as explained below, is primarily an over-sampling technique \citep{Krawczyk2016}.

The motivation for using an augmentation technique such as SMOGN is twofold. First, we want to be
able to reproduce underrepresented data that are valuable in the context of characterising the
halo--galaxy connection. Second, we are interested in boosting our statistics in order to measure
the clustering properties of TNG300 galaxies. 
Specifically, SMOGN works by combining random under-sampling with two over-sampling techniques:
SmoteR (an adaption for regression of Smote\footnote{Synthetic Minority Over-sampling Technique (Smote) is a data augmentation method that improves imbalanced classification data sets. It works by over-sampling some minority class and under-sampling the majority class, using kNN. The main difference between Smote and SMOGN is that the first is built to predict discrete results, while the second is able to provide continuous predictions.}
\citep{Chawla_2002}) and Gaussian noise. In essence, the algorithm first bins the data for a
given target variable and subsequently splits the resulting distribution in ``rare'' and
``normal'' bins. Rare bins are augmented, whereas normal bins are under-sampled. 
In the case of augmentation, an object/datum within a rare bin is selected, and their $k$ nearest neighbours are determined by measuring their Euclidean distances to the initial object. Subsequently, a neighbour is randomly chosen. If this neighbour is close enough to the the initial object, a new object is generated by interpolating between them. This is a ``safe distance" (following the nomenclature of \citealt{pmlr-v74-branco17a}), defined for the neighbours; this distance is based on a simple threshold, i.e., half the median distance of all $k$ neighbours. Neighbours within the safe distance are picked, whereas those outside are not. During this over-sampling process, Gaussian noise creates new objects closer to the original sample. Under-sampling works in a similar way, by randomly removing objects from the original bins.
The final product is a distribution with a higher number of objects in the previously underrepresented region and a lower number of objects in the region around the bulk of the distribution. In this work we use the \href{https://pypi.org/project/smogn/}{\texttt{smogn}} library \citep{Kunz2019}. The generated distributions are provided in the Appendix \ref{appendix}.

\begin{figure*}
    \centering
    \includegraphics[scale=0.45]{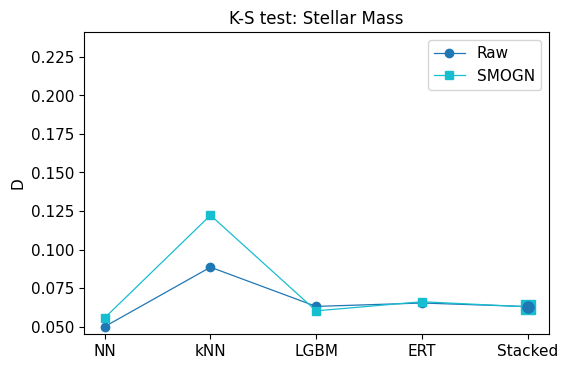}
    \includegraphics[scale=0.45]{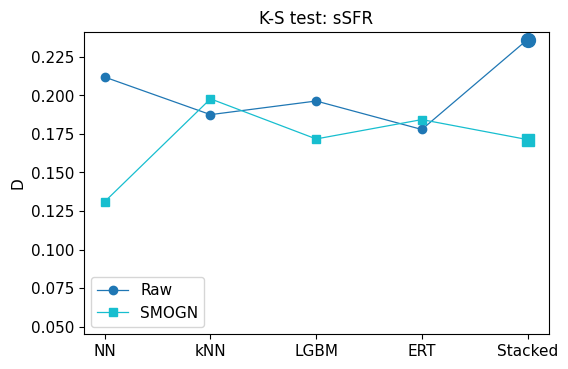}
    \includegraphics[scale=0.45]{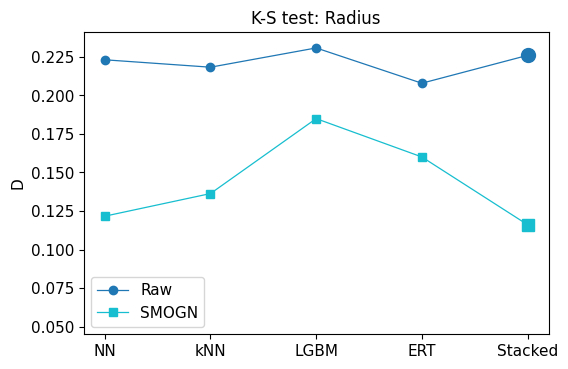}
    \includegraphics[scale=0.45]{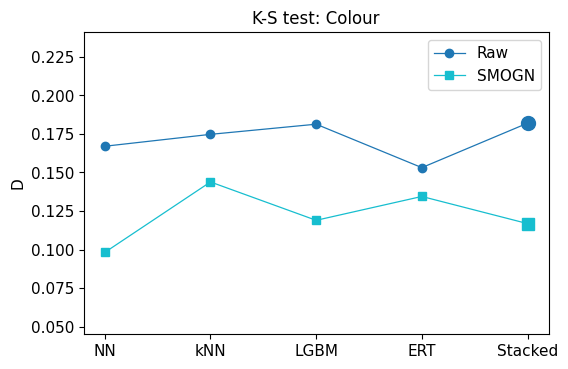}
    \caption{K-S test between the true and the predicted distributions, as described in Eq. \eqref{eq:ks-test}, for all galaxy properties and algorithms. Lower values correspond to better fits. Note that we have opted to connect the dots despite the fact that the properties in the x-axis are not correlated. This format is employed in this and other subsequent plots in order to facilitate the readability of the plots.}
    \label{fig:KS-test}
\end{figure*}

\subsection{Loss functions and performance metrics}\label{metrics}

As mentioned above, the metrics provide a mean of measuring the difference between network
predictions $y^{\mathrm{pred}}$ and the target values $y^{\mathrm{true}}$ (from either the
validation or test set). When used at the training stage, they are usually called loss functions
(in the case of NN and LGBM) or simply metrics otherwise. They are generally determined by the
class of the ML problem, the following being the most general choices for regression (and also
used in this work): 

\begin{itemize}
\item {\bf Mean Squared Error (MSE):}
   \begin{equation}
       \text{MSE} = 
       \frac{1}{n} 
       \sum_{i = 1}^n \left( y^{\rm{pred}}_i - y^{\rm{true}}_i \right)^2
   \end{equation}

\item {\bf Pearson Correlation Coefficient (PCC):}

 \begin{equation}
 \label{eq:PC}
 \rm{PCC} = 
       \frac{\rm{cov} \left( y^{\rm{pred}}, y^{\rm{true}} \right)}
       {\sigma_{y^{\rm{pred}}} \sigma_{y^{\rm{true}}}}
 \end{equation}

\end{itemize}

We have used MSE to monitor the underfitting/overfitting trade-off in LGBM and NN, as a loss function. The PCC score was used as an additional test, as another indicator for the performance of the methods, in the test set, for all the methods.

Another useful performance evaluation method in the context of ML (and other statistical
analyses) is the well-known Kolmogorov-Smirnov test (hereafter K-S test), which quantifies the 
difference between two distributions in a non-parametric way (see, e.g., \citealt{ivezic2014statistics}). In essence, the K-S test measures the maximum distance between
the two cumulative distributions ($F_i (x_1)$, $F_i (x_2)$), namely: 
\begin{equation}
    D = \text{max} \left( |F_1 (x_1) - F_2 (x_2)| \right). \label{eq:ks-test}
\end{equation}
When considering not only one, but two independent variables, it may also be useful to compute the 2D K-S test. The method is essentially the same as for the 1D K-S test, but accounting for two-dimensional data. This algorithm, developed mostly for astronomical analyses, compute the cumulative distributions along the coordinate axes of the two variables. 
More details can be found in \citealt{2DKS-Peacock1983, 2DKS-Fasano1987}. In this work we have used the \cite{2DKS} repository.

\begin{figure}
    \centering
    \includegraphics[scale=0.37]{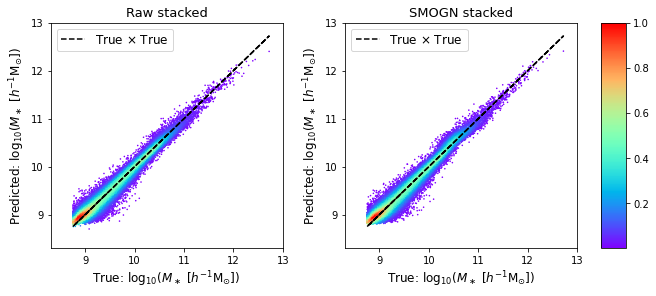}
    \includegraphics[scale=0.37]{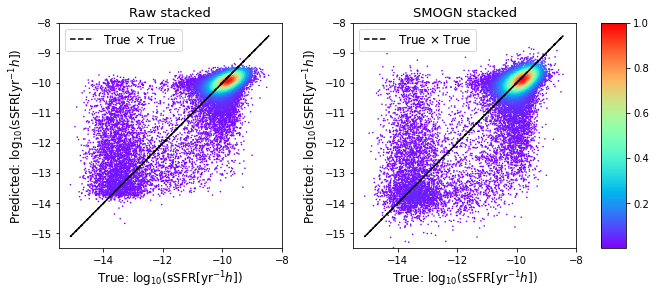}
    \includegraphics[scale=0.37]{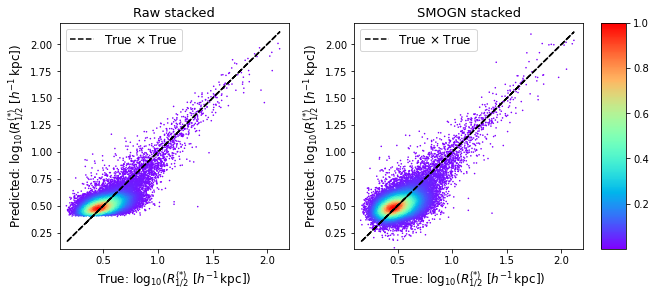}
    \includegraphics[scale=0.37]{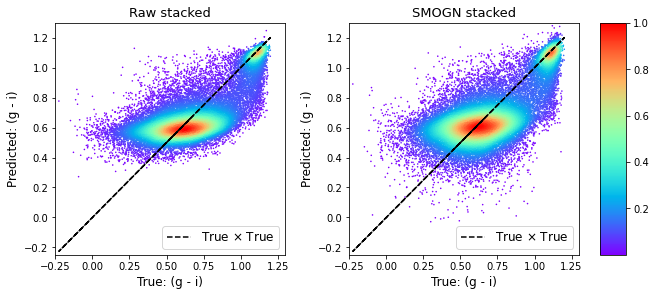}
    \caption{Scatter plots for the predicted {\em v.} true distributions for stellar mass, sSFR, radius, and colour, for raw (left) and SMOGN (right) stacked models. The colour code indicate the normalized density of objects.}
    \label{fig:predXreal}
\end{figure}

\section{Results}
\label{results}

In this section, we present the main results of our analysis, both in terms of our galaxy-property predictions and our clustering measurements. 

There are multiple ways in which the extent to which our methods are able to recover the central galaxy properties can be measured. First, in Section \ref{sec:pred_and_perf} we compare the true and predicted distributions of those properties, verifying to what extent the ML method is able not only to recover the main populations, but also the outlying, less frequent objects that reflect the diversity in the galaxy properties. This comparison can be seen directly in the distributions, and it can also be expressed in terms of summary statistics such as the PCC and the K-S test. 

The next step is to understand the relative relevance of the (input) halo properties in terms of the galaxy predictions. More precisely, this {\it feature importance}, which is presented in Section \ref{sec:FI}, indicates which halo property matters more for each type of prediction, and allows us to determine the driving factors behind those properties. This part of the analysis is, of course, particularly relevant in the context of the physical meaning of our predictions. The differences between the feature importance obtained from different methods is also indicative of the complementarity between those methods, which is a key motivation for the stacking method.

Finally, in Section \ref{sec:PS}, we show a key diagnostic of the success and failures of the ML predictions, which is provided by the power spectra of the resulting galaxy populations, when the sample is split in terms of those properties. Although the power spectra are rather complex summary statistics for the accuracy of the predictions, they are of key significance for the cosmological applications of those methods. 

\subsection{Predictions and performance}
\label{sec:pred_and_perf}

Our main goal is to recover some of the main galaxy properties, which means,  in particular, to reproduce the frequencies (distributions) of each property. The true and predicted distributions are shown in Fig. \ref{fig:histograms} for stellar mass (first row of plots), sSFR (second row), radius (third row), and colour (fourth row).
The true distributions are shown as the filled regions, while the distributions for the predictions using the ML methods are shown as lines.
The first column shows the results for the ``raw'' ML models, which employ the original training set drawn from the TNG300 catalogue.
The second column corresponds to the SMOGN models, i.e., the ML models trained with the data which is augmented using the SMOGN technique. Finally, the third column shows the distributions for the stacked models (using all the different individual ML models) from the raw and SMOGN data sets. 

The plots in Fig. \ref{fig:histograms} clearly show that our machinery is capable of recovering
the general distributions. As expected, stellar mass is the property that is better predicted by
the models, with only some small deviations in the distributions. For the rest of the properties,
the performance of the machine worsens, indicating a more complex connection with the properties
of the hosting haloes (potentially due to information that we are not taking into account in our
input sample). Despite these problems, the method is able to reproduce the main features of the
distributions, i.e., the general form of the distribution for galaxy size, and the bimodality for
colour and sSFR. The different individual algorithms provide similar results,
mainly in the case of the raw models, but less so in the case of the SMOGN results, where the predicted distributions indicate that the tree methods still tend to privilege the peaks. This results in smaller improvements in the predictions at the tails of the distributions. However, very large
deviations are found when the raw, augmented, and stacked models are compared. These differences
are highlighted by the stacking ensemble models, which combine all the different ML models. 
The raw models are slightly less efficient at reproducing the scatter in the galaxy properties,
especially towards the tails of the distributions, where underrepresented populations lie. 
They are, however, better at recovering the regions around the peaks of the distributions. 
Although the SMOGN methods are better at recovering the overall shapes of the distributions, we also 
notice the appearance of some artefacts, such as the small hump in the number of predicted
galaxies around $\log_{10} (M_{\star} [h^{- 1} M_{\odot}]) = 10.5$, which may be due to the SMOGN binning choices to under-sample low-mass objects and to over-sample large-mass ones (see the first histogram of Appendix \ref{appendix}, Fig. \ref{fig:appendix_SMOGN}).

\begin{figure}
    \centering
    \includegraphics[scale=0.37]{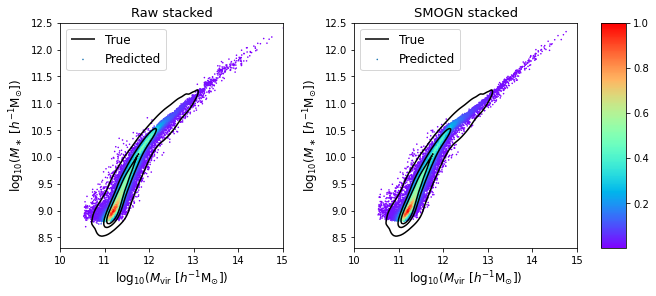}
    \includegraphics[scale=0.37]{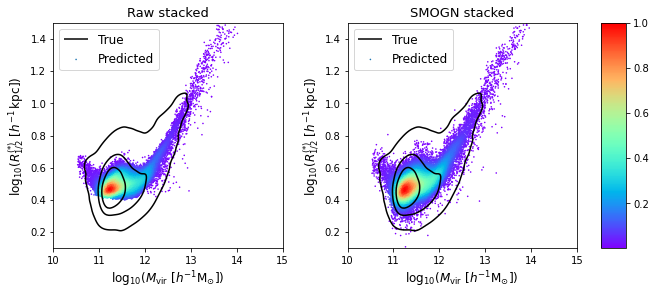}
    \includegraphics[scale=0.37]{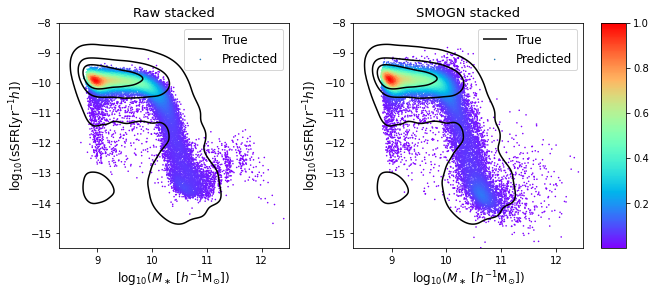}
    \includegraphics[scale=0.37]{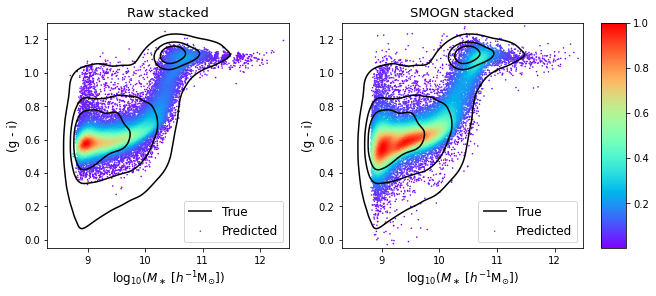}
    \includegraphics[scale=0.37]{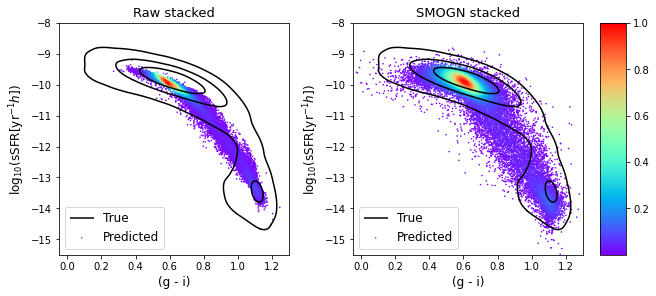}
    \caption{Joint galaxy properties: from top to bottom, stellar mass {\em v.} halo mass, radius {\em v.} halo mass, sSFR  {\em v.} stellar mass, colour  {\em v.} stellar mass, and sSFR  {\em v.} colour. For all relations, the stacked raw (left) and SMOGN (right) models are compared. 
    The colour code represents the normalized density of objects and the
    true distributions are shown in black contours.}
    \label{fig:astrophysical_plots}
\end{figure}

Although a visual inspection of the overall shapes of the distributions is already indicative of
the performance of the methods, the metrics of Section \ref{metrics} provide a quantification of
the deviations. In Fig. \ref{fig:model_score}, we show the PCC
of Eq. \eqref{eq:PC}, for which a value of 1 corresponds to a perfect match. 
This figure confirms that the best predicted galaxy property is stellar mass, both in terms of
the raw, the SMOGN and the stacked models, reaching values of $\sim 0.98$. The Pearson
correlation coefficient drops to $\sim 0.80$ for sSFR, $\sim 0.7-0.77$ for radius, and 
$\sim 0.57-0.71$ for colour. The raw and SMOGN models perform, in general, similarly. 
The main difference resides in the lower scores for kNN and NN in the SMOGN models, particularly
for colour. This slightly worse performance of the SMOGN models for some of the galaxy properties
is directly connected to the augmentation technique itself. As mentioned before, the SMOGN method
tends, by construction, to give greater weights to the tails of the distributions, especially in
the case of colour and radius. 
Together with the PCC results, we present in Appendix \ref{appendix_scores}, Table \ref{tab:score_comp}, the exact values for MSE and PCC, measured in the test subset for all galaxy properties, using the raw and SMOGN stacked models.

\begin{figure*}
    \centering
    \includegraphics[scale=0.45]{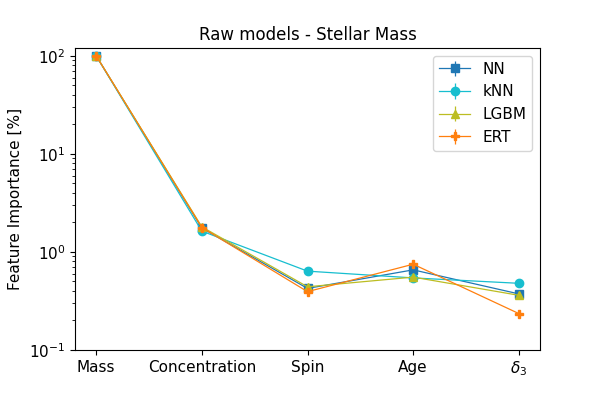}
    \includegraphics[scale=0.45]{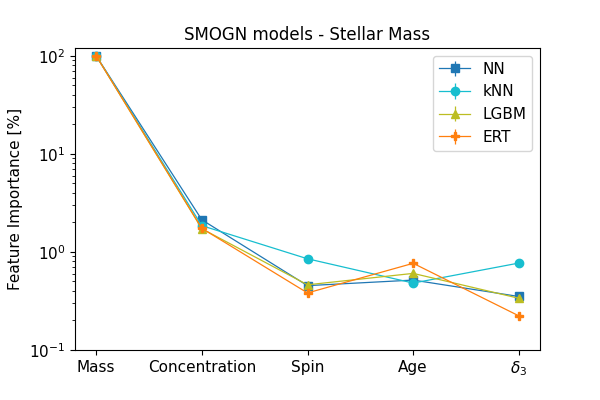}
    \includegraphics[scale=0.45]{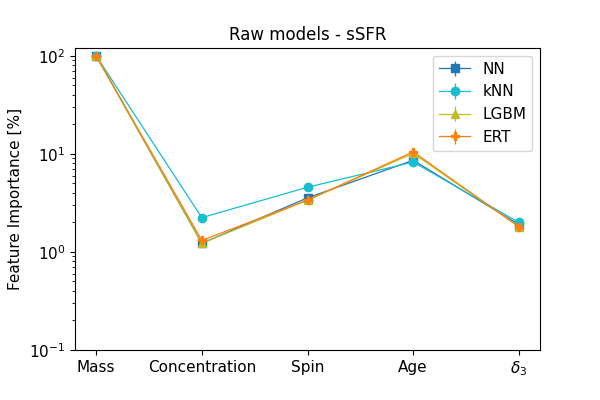}
    \includegraphics[scale=0.45]{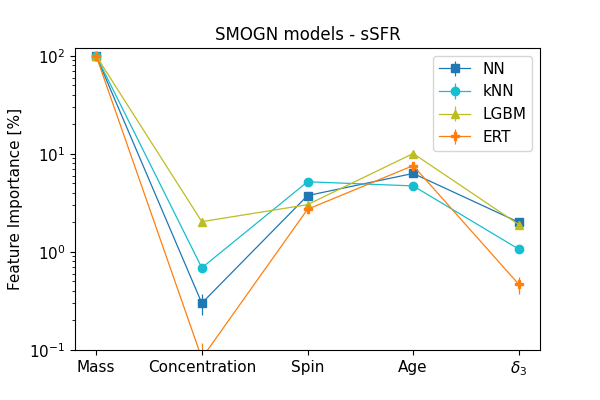}
    \includegraphics[scale=0.45]{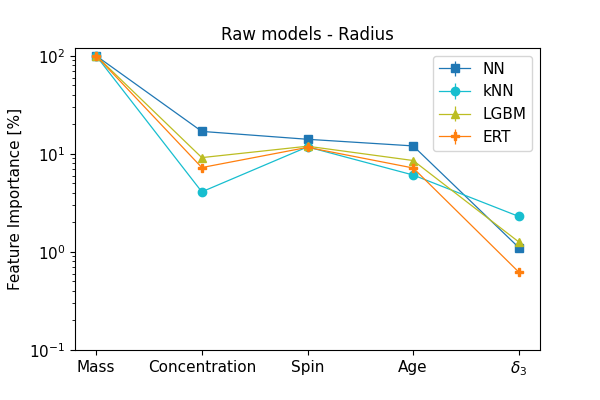}
    \includegraphics[scale=0.45]{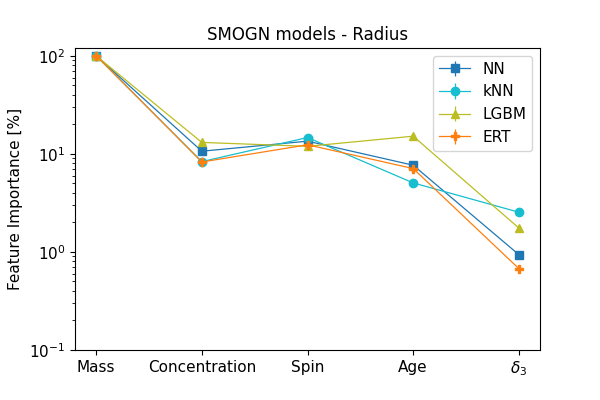}
    \includegraphics[scale=0.45]{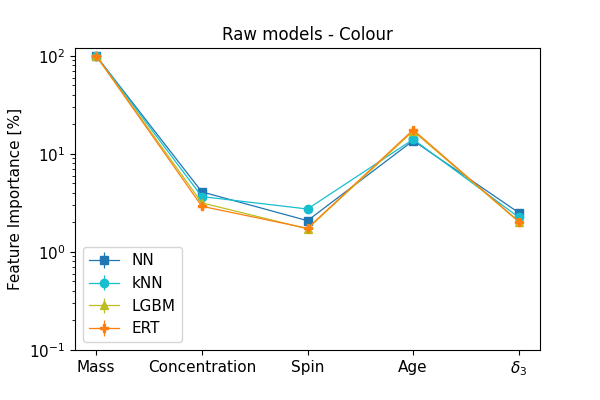}
    \includegraphics[scale=0.45]{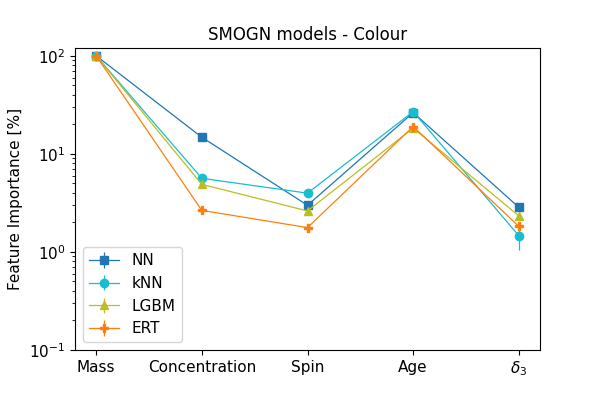}

    \caption{Permutation feature importance, computed using MSE as a score, for raw models (left) and
    SMOGN models (right). Each row corresponds to the predictions for stellar mass, sSFR, radius, and
    colour, respectively. This plot shows the weights given to each individual halo property (mass,
    concentration, spin, age and $\delta_3$), normalized with respect to the feature importance of the
    halo mass parameter.}
    \label{fig:feature_importance}
\end{figure*}

Another summary statistics that can serve as a metric to compare the true and predicted
distributions is the K-S test, which is shown in Fig. \ref{fig:KS-test}. In this test, values
closer to zero denote that the distribution of predicted values resembles better the true
distribution. 
From Fig. \ref{fig:KS-test}, it is again clear that stellar mass is the most easily predicted
galaxy property, while the remaining properties are harder to determine just on the basis of the
halo properties. Interestingly, the results of the K-S tests also indicate that the distributions
of predicted radii and colours (and even sSFRs) for SMOGN reproduce better the true
distributions. This result, again, reflects the philosophy behind the SMOGN technique. 
Finally, Fig. \ref{fig:KS-test} clearly motivates the use of the stacked models 
(represented by the big markers on the right-hand side of each panel), as their performances are 
often very good in terms of recovering the true distributions. 
These results show that the stacked models are capable of providing a fair combination of the predictions of the different models.

\begin{figure}
    \centering
    \includegraphics[scale=0.45]{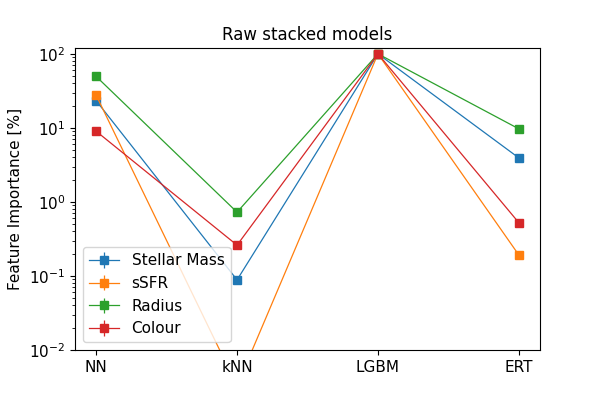}
    \includegraphics[scale=0.45]{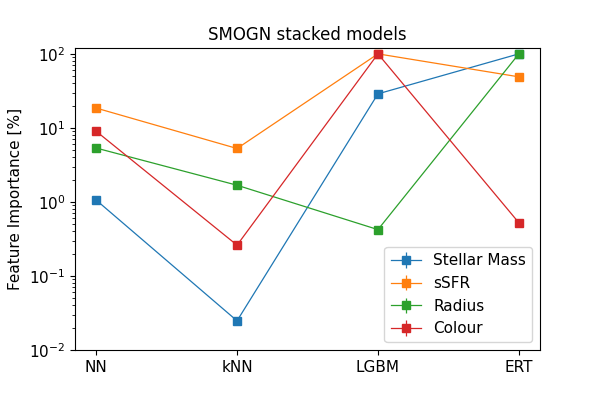}
    \caption{Permutation feature importance computed for the stacked raw (upper panel) and SMOGN (lower panel) models, for each of the predicted galaxy properties (stellar mass, sSFR, radius, and colour). This plot shows the weights given to each individual ML model (NN, kNN, LGBM, and ERT) in the stacking procedure, normalized with respect to the feature importance of the LGBM model.}
    \label{fig:feature_importance-stacked}
\end{figure}

Once we have addressed the distributions of galaxy properties, we proceed to analyse the
predictions on a one-to-one basis. Fig. \ref{fig:predXreal} displays the scatter plots of the
true {\em v.} the predicted values, for 30,000 galaxies randomly chosen from the test sample. 
The colour code represents the normalized density of objects. From these plots, it remains clear
that the galaxy property that displays the most direct connection with the halo properties
considered is stellar mass, as the small scatter in the upper panels of Fig. \ref{fig:predXreal}
demonstrates. Our predictions become more uncertain for the rest of the galaxy properties. 
For sSFR, the models, particularly the SMOGN-augmented ones, perform relatively well towards the
bulk of the distribution. However, despite our attempts to palliate the effect of the null-SFR
values, objects with very small sSFR are still problematic for the machine. In the case of galaxy
radius, our predictions for the largest objects are good and unbiased. For the smaller, more
common objects, the raw model predicts a distribution that is tilted with respect to the real one. 
This effect is due to the fact that the machine predicts a narrower range of values for this property.
Something similar happens for galaxy colour, where, again, the bimodality is well reproduced, but
the predicted blue cloud is severely tilted as compared to the real data (an effect that is not
as strong for the red sequence). 
A very important advantage of SMOGN seems to emerge here. As shown in the right-hand column of Fig. \ref{fig:predXreal}, SMOGN tends to rectify this problem, reducing the tilt in the distributions. This improvement, which is still not complete, does suggest that using the augmented data set allows the machine to predict a wider range of output values. The effect is particularly evident for colour and radius, where the raw models are unable to predict any galaxies with (g - i) < 0.3 and radii lower than 0.375 $h^{-1}$ Mpc, whereas the SMOGN results do. This is, again, an encouraging advantage of the SMOGN models that will be explored in more depth in the future.

Finally, it is important to analyse the ability of the method to reproduce the relations between different galaxy and halo properties. Fig. \ref{fig:astrophysical_plots} shows, from top to bottom, stellar mass {\em v.} halo mass, radius {\em v.} halo mass, sSFR {\em v.} stellar mass, colour {\em v.} stellar mass, and sSFR {\em v.} colour (in the same format of Fig. \ref{fig:predXreal}). Generally speaking, Fig. \ref{fig:astrophysical_plots} displays encouraging results, demonstrating that the machine produces fairly realistic relations between properties. One of the main problems to overcome, as this figure illustrates, is the scatter in these relations. By construction, ML models tend to concentrate on the bulk of the distributions, which hinders the prediction of scatter. A clear example of this is the mass--size relation in the second row of Fig. \ref{fig:astrophysical_plots}. Again, SMOGN works in the right direction, increasing the scatter in the relations. Further investigation is still needed in order to reproduce the scatter in these joint galaxy relations.

\begin{table}
 \caption{\label{tab:2DKS} 2D K-S test for the stacked raw and SMOGN models for the joint galaxy distributions.}
 \begin{center}
  \begin{tabular}{ccc}
   \cline{1-3}
   \textbf{Joint property} & \textbf{D (raw stacked)} & \textbf{D (SMOGN stacked)} \\
   \cline{1-3}
    Stellar Mass {\em v.} Halo Mass & $0.065683$ & $0.065417$\\
    Radius {\em v.} Halo Mass & $0.228733$ & $0.116850$\\
    sSFR  {\em v.} Stellar Mass & $0.272300$ & $0.208417$\\
    Colour  {\em v.} Stellar Mass & $0.224533$ & $0.164350$\\
    sSFR  {\em v.} Colour & $0.327667$ & $0.282467$\\
   \cline{1-3}
   \end{tabular}
  \end{center}
\end{table}

In order to quantify the results of Fig. \ref{fig:astrophysical_plots}, we have computed the 2D K-S test between the true and predicted joint distributions. The results of this test are presented in Table \ref{tab:2DKS}. For all pairs of properties, the distances between the cumulative distributions are close to zero, and here again the best result is for stellar mass {\em v.} halo mass. The decrease of $D$ value for each relation from raw to SMOGN models is remarkable, reaching its highest difference for radius {\em v.} halo mass, followed by colour  {\em v.} stellar mass.

\begin{figure*}
    \centering
    \includegraphics[scale=0.582]{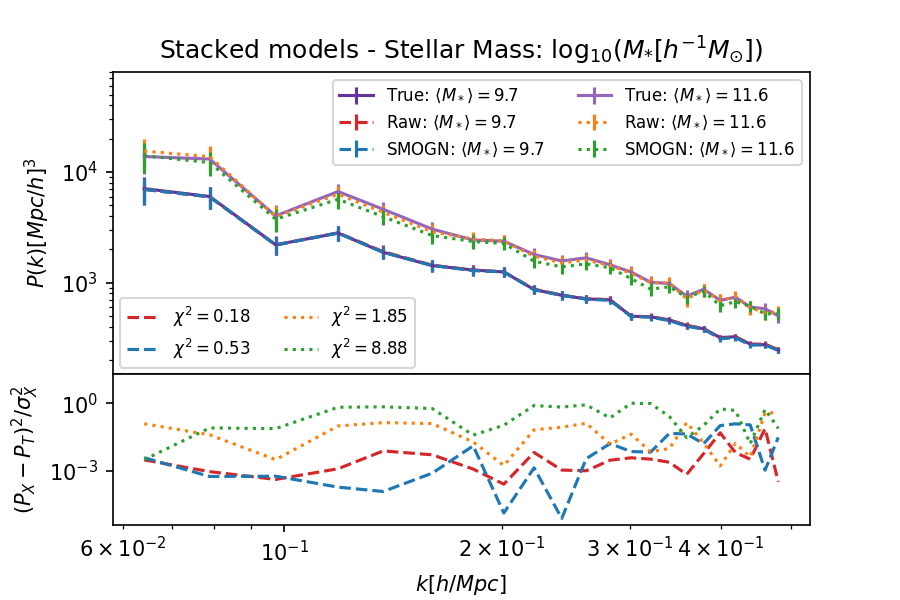}
    \includegraphics[scale=0.582]{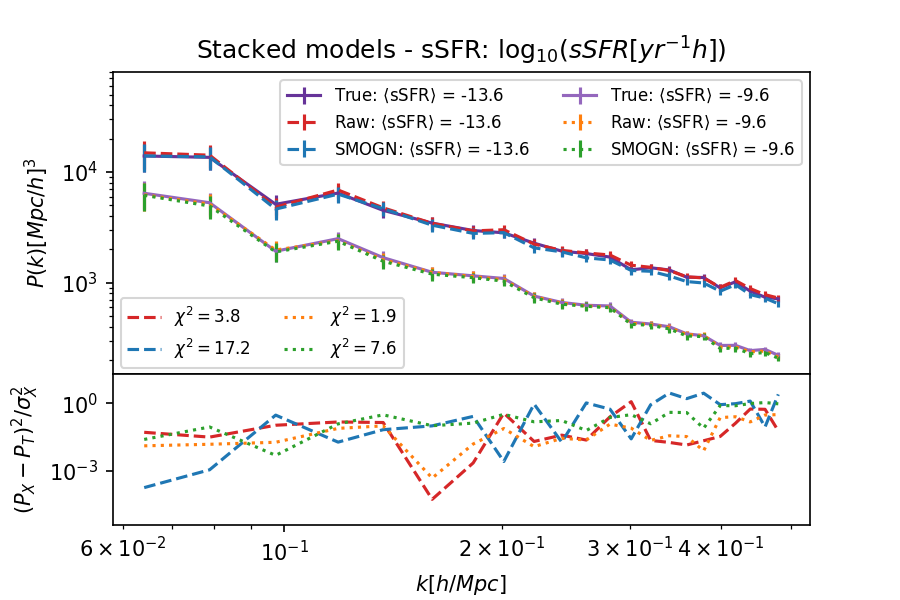}
    \includegraphics[scale=0.582]{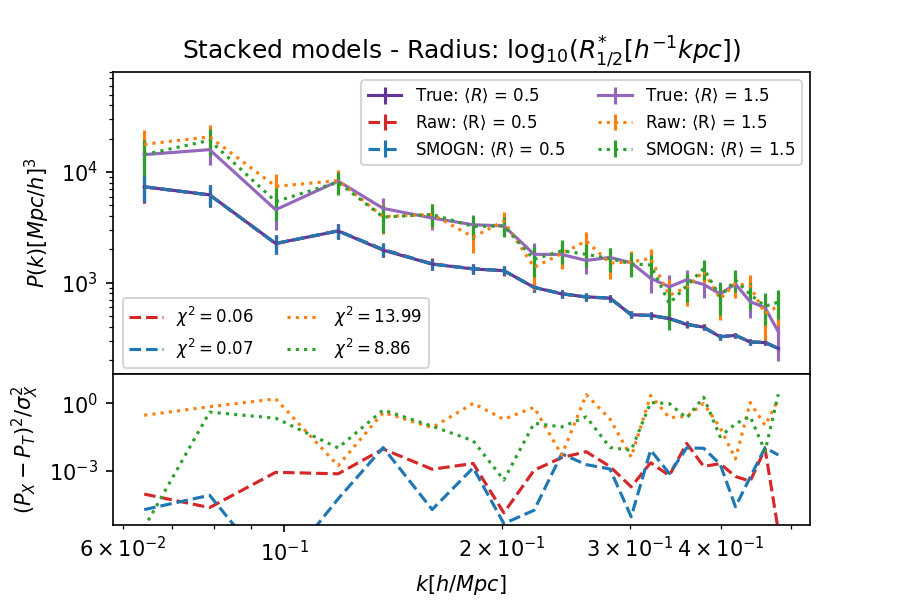}
    \includegraphics[scale=0.582]{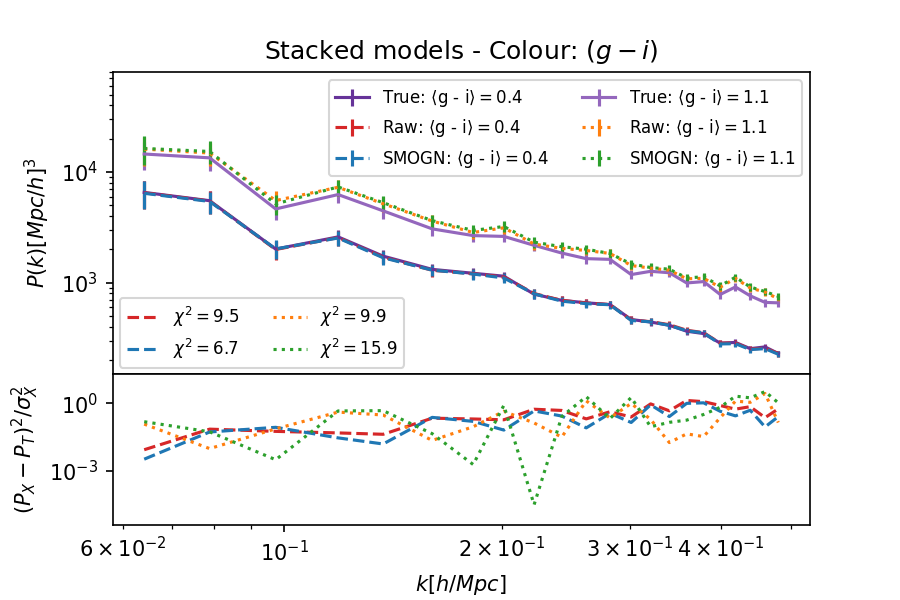}
    \caption{The power spectra for the raw and SMOGN stacked models compared with the true data set. In each panel, a different galaxy property, split in two bins, is analysed (see text). The residuals are provided in the subplots.}
    \label{fig:power_spectrum}
\end{figure*}

\subsection{Feature Importance Analysis}
\label{sec:FI}

The ultimate goal of our analysis is to establish relations that allow us to shed light onto the intricacies of the halo--galaxy connection. One of the ways to address this aspect and to gain some insight into the inner workings of the ML methods is to analyse the weights given to each feature (i.e., halo property) which contributes to producing the desired output.

There are different methods to compute this {\it{feature importance}} (some methods, such as the tree-based methods, even come with a built-in machinery to compute this statistic).
In order to compare the relative weights of the input features for the different ML methods, we have chosen to compute the so-called permutation feature importance, which is basically the relative decrease in a particular score (in our case, the MSE), measured for different runs of the model after randomly removing one feature at a time (\citealt{breiman2001})\footnote{We use  \href{https://eli5.readthedocs.io/en/latest/blackbox/permutation_importance.html?highlight=PermutationImportance}{\texttt{eli5}} to perform this computation.}. For clarity, all our results have been converted to percentage of the feature importance for halo mass, which is the predominant parameter for determining all galaxy properties.

Fig. \ref{fig:feature_importance} shows the importance of the halo features for each individual ML model. Here, a hierarchy emerges: on the one hand, the halo mass is clearly the most important feature for the prediction of all the galaxy properties, as expected. On the other hand, the environmental property $\delta_3$ had an almost
negligible level of importance for all the galaxy properties. 
Somewhere in the middle, age turns out to be quite important for colour and sSFR, but less so for stellar mass and radius.
Galaxy radius is perhaps the most interesting case, where a number of halo features appear in a less hierarchical way, with concentration, spin, and age contributing at about the same level ($\sim 10\%$ compared to halo mass).

\subsubsection{Model importance analysis applied to the stacked models}

We can also apply the idea of feature importance to the stacked models, measuring the level of importance of the different ML algorithms.
In that case, the features under analysis are the predictions of the individual ML models. The results have been normalized to the LGBM predictions, since that is the predominant model for determining the stacked predictions for all galaxy properties (the same way that halo mass was the predominant feature before).
The results are shown in Fig. \ref{fig:feature_importance-stacked}, where the upper and lower 
panels correspond to the raw and SMOGN models, respectively.
Recall that, for the raw models, the best predictions are obtained using the LGBM and NN methods, while the kNN and ERT methods return worse predictions, depending on the galaxy property. The SMOGN methods behave differently, with LGBM and ERT yielding the best results.

\subsection{Power Spectrum}
\label{sec:PS}

An important test to our methods is the clustering properties of the galaxies whose properties we are trying to predict. By splitting those galaxies in two populations according to those properties, and then computing the clustering of each population, we can check whether our predictions are able to separate the galaxies correctly, according to the types of haloes that they inhabit. Furthermore, given that we have exactly the same dark matter haloes, by splitting the galaxy populations both in terms of their predicted properties, as well as their true properties, we can assess some of the systematics that arise in the bias of those populations as a result of our imperfect predictions, in a way that is protected against cosmic variance.

\begin{table}
 \caption{\label{tab:PS_edges_and_means} Bin edges and central values for the subsets considered for the computation of the power spectrum.}
 \begin{center}
  \begin{tabular}{ccc}
   \cline{1-3}
   \textbf{Property} & \textbf{Bin edges} & \textbf{Central values} \\
   \cline{1-3}
    $\log_{10} ( M_\ast{}$ [$h^{-1} {\rm M_{\odot}}$] ) & $[8.8, 10.6, 12.7]$ & $[9.7, 11.6]$\\
    $\log_{10} ( {\rm sSFR} [ {\rm yr^{-1}} h] )$ & $[- 16.3, - 11.0, - 8.3]$ & $[- 13.6, - 9.6]$\\
    $\log_{10} ( R_{1/2}^{(*)}$ [$h^{-1} \, {\rm kpc}] )$ & $[0.1, 0.9, 2.1]$ & $[0.5, 1.5]$\\
    $(g - i)$ & $[- 0.23, 1.0, 1.2]$ & $[0.4, 1.1]$\\
   \cline{1-3}
   \end{tabular}
  \end{center}
\end{table}

We have split the galaxies according to each property (stellar mass, sSFR, radius and colour) in two bins each, with bin edges and central values listed in Table \ref{tab:PS_edges_and_means}. For the true galaxies, we use their positions from the TNG300 catalogue, while for the ML predictions we use the positions of their hosting haloes. All the spectra were then measured for the entire TNG300 box, using the Python package \href{https://nbodykit.readthedocs.io/en/latest/}{\texttt{nbodykit}} \citep{nbodykit2018}. Because we have only one single IllustrisTNG box, the uncertainties of the spectrum on each of the Fourier bins (bandpowers) $k_i$, for each tracer $\alpha$, $\sigma_{P_{\alpha, i}}$, were assumed to be given by:
\begin{align}
  \frac{\sigma^2_{P_{\alpha, i}}}{P_{\alpha, i}^2} = \frac{2}{\tilde{V}_i V} \, \left(
  \frac{1 + \bar{n}_{\alpha} P_{\alpha, i}}{\bar{n}_{\alpha} P_{\alpha, i}}
  \right)^2 \; ,
\end{align}
where $ \tilde{V}_i = 4 \pi k_i^2 \Delta k/(2 \pi)^3 $ is the volume of the Fourier bin, with $\Delta k$ representing the width of the bandpower, $V$ is the volume of the catalogue, and $\bar{n}_{\alpha}$ is the mean number density of the tracer $\alpha$ (which here stands for the two bins in galaxy properties). 

The results for the power spectrum are shown in the four panels of Fig. \ref{fig:power_spectrum}, which, from upper left to lower right, correspond respectively to binning the galaxies in stellar mass, sSFR, radius, and colour. The legends in the upper right corner show the central values
of the bins of each galaxy property, for each bin and method -- which, as can be seen, are identical for all methods. The legends in the lower left corners indicate the $\chi^2$ values for the fit of the spectra of the predicted versus the true galaxies. 

For each plot in Fig. \ref{fig:power_spectrum} we also show the residuals:
\begin{align}
   \frac{\left[ P_{\alpha, i}^\textit{Pred} (k) - P_{\alpha, i}^\textit{True} (k) \right]^2}{\left[ 
   \sigma_{P_{\alpha, i}}^\textit{Pred}
   \right]^2} \; , \label{eq:redisual}
\end{align}
where $Pred$ is the predicted and $True$ is the true power spectrum (i.e., the power spectrum that results from using the original galaxy property and position in the TNG300 box). For stellar mass, the residuals range between
$10^{- 5}$ and 0.12, for the less massive bin, and $1.6 \cdot 10^{- 3}$ and 1 for the most massive interval. In the case of sSFR, the residuals range between $5.5 \cdot 10^{- 5}$ and $2.9$ for the first (low sSFR) bin, and 
$5 \cdot 10^{- 4} - 1.0$ for the second subset (high sSFR). Similar results are obtained for galaxy size and colour. For the former, ranges of 
$[2 \cdot 10^{- 8}, 0.02]$ and $[3 \cdot 10^{- 6}, 2.5]$ are obtained for the residuals in each bin, respectively. Finally, the residuals in the power spectrum for galaxy colour are $[0.009, 1.3]$ and $[3 \cdot 10^{- 5}, 3.3]$, for the blue and red subsets, respectively. 

Particularly, the residuals follow the same trend for sSFR and colour, and behaves differently for
stellar mass and radius. 
In the former case, the comparison happens because both bins either have their mean values (for the raw and SMOGN predictions) close to the true spectra, or because the dispersion $\sigma$ is higher enough (specifically in the case of the red colour bin). 
In the latter case, it is evident that the bins with a low number of objects (higher stellar masses and higher radii) have higher values for the residuals.
Some trends, such as the fact that the residuals increase with $k$ for all four predicted properties, show that our predictions are more accurate on larger scales. This can be both because of shot noise (which affects more the small scales), and also because it is harder for the predictions to match precisely the local environments of those galaxies and halos. 
Another point to consider is that binning the galaxy populations sometimes leads to  samples with very different sizes, which also affects the residuals.

When adding up the residual for all the values of $k$, we obtain the $\chi^2$ associated to each power spectrum. Since we have 22 bins of $k$, the $\chi^2$ per degree of freedom is significantly smaller than 1 in all cases, which is indicative of an excellent agreement. It is noteworthy that the SMOGN stacked models perform slightly worse for stellar mass and sSFR, compared with the raw stacked models, but they do better for radius and colour.

\section{Discussion and conclusions}
\label{discussion}

The predictive power of ML techniques can be harnessed to reproduce the hidden intricacies of the halo--galaxy connections. The main goal in this field is to establish relations between the properties of galaxies and the properties of their hosting haloes, in the cosmological context of the LSS of the Universe. This problem can be treated in ML in terms of an {\it{input}} data set (halo properties), which is known a priori, and an {\it{output}} data set, corresponding to the galaxy properties that we attempt to predict (\citealt{Kamdar2016, Agarwal2018, Calderon2019, Jo2019, Man2019, Kasmanoff2020, Lovell2021, shao2021}). 

We have selected four different ML algorithms (NN, kNN, LGBM, ERT), as well as the combination of their predictions (the stacked models), and evaluated their ability to predict stellar mass, colour, sSFR, and half-mass radius for central galaxies in the TNG300 hydrodynamical simulation.
In addiction, we have employed a data augmentation technique called SMOGN for the first time in the context of the halo--galaxy connection field.
Our set of halo properties includes halo mass, age, concentration, spin, and overdensity around
haloes.
Overall, our findings are consistent with previous results in the literature, with stellar mass being the most accurately predicted property, with a PCC of $\sim 0.98$ (previously reported values are 
typically $0.92 - 0.957$, see \citealt{Kamdar2016, Agarwal2018, Lovell2021}). 
The second best-predicted property is sSFR, with a correlation coefficient of $\sim 0.8$ 
(previously, $0.745 - 0.794$, see \citealt{Kamdar2016, Agarwal2018, Lovell2021}). 
For size and colour we obtain coefficients in the range $0.7-0.8$ and $0.59-0.71$, respectively. 
A similar hierarchy for the predictive power of our ML methods is obtained when other performance
estimators such as the K-S test are employed.

\begin{figure*}
    \centering
    \includegraphics[scale=0.4]{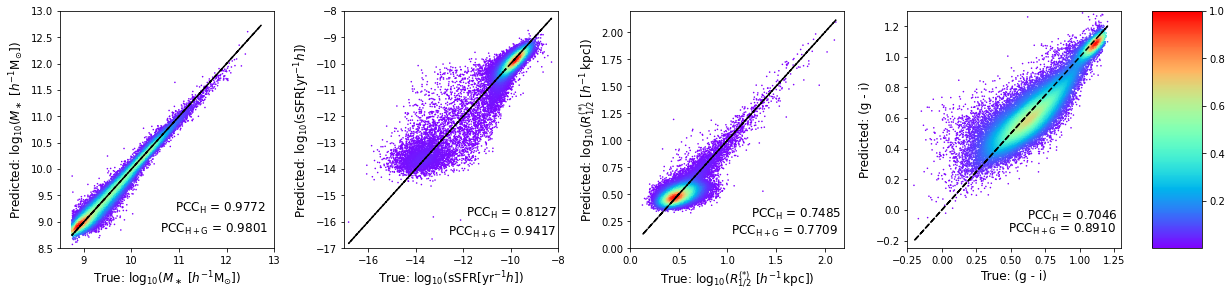}
    \caption{Scatter plots of the predicted {\em v.} true distributions for stellar mass, sSFR, radius, and colour for NN models produced using all halo and galaxy properties (except for the one under analysis). Each point represents a central galaxy and the colour bar corresponds to the normalized density of objects in each region.}
    \label{fig:including_gal_props}
\end{figure*}

The aforementioned improvements are primarily due to the use of both the stacked models and the SMOGN technique. The stacked models perform a linear combination of the different ML predictions for each galaxy property. The SMOGN method, on the other hand, alleviates the problem of imbalanced data sets, by statistically compensating the lower-populated regions in parameter space. Note that ML methods tend to focus, by construction, on reproducing the better represented data. Needless to say that, in the context of the halo--galaxy connection, there is significant value in the sparse data (rare objects). We have shown in a quantitative manner that the use of SMOGN has several advantages. First, it helps predicting galaxies in the tails of the distributions (this can be seen from the distribution histogram or from the small D-values obtained from a K-S test: e.g., $D \lesssim 0.175$ for sSFR and $D \lesssim 0.125$ for radius and colour). Second, it tends to rectify the tilted real v. predicted distributions obtained using the raw, unaugmented models, by expanding the predicted range of values.

The above advantages are also noticeable when the joint distributions of galaxy/halo properties are analyzed. We have demonstrated that we are capable of reproducing well the shape of several important relations, such as the stellar/halo mass relation (SHMR) or the galaxy size -- halo mass relation, to name but a few. Here, SMOGN proves once again to be of help, particularly in terms of reproducing the overall scatter in the relations. We have used the 2D K-S test to quantify the accuracy of our predictions, both for the raw and for the SMOGN stacked models. To give an example, the use of the data augmentation technique improves our predictions by $\sim 49$ and $\sim 27\%$ (as compared to the raw models) for the very relevant galaxy radius -- halo mass relation and the colour--stellar mass relation, respectively. In conclusion, the results presented in this paper clearly show the potentialities of SMOGN in the context of the halo--galaxy connection.

In terms of the physical implications of our results, the most important aspect is the quantification of the {\em feature importance}, i.e., the contribution of each halo property to the prediction of each galaxy property. 
The validity of this analysis is of course due to the high consistency across the different models. 
As expected, stellar mass seems to be almost completely determined by halo mass, whereas the inclusion of halo age is necessary to predict both sSFR and colour. 
Maybe the most interesting case is again galaxy size, which, within the uncertainties of our analysis, seems to be primarily determined by halo mass (this makes sense due to the mass--size relation). 
The contribution of other properties such as spin, concentration or age seem to be equally relevant and significant, but it is important to bear in mind that our prediction for galaxy radius is still not optimal. These results seem to be connected with the shape of the (halo/stellar) mass--size relation (see, e.g., \citealt{Rodriguez2021}): at the high-mass end, central galaxy size is proportional to halo mass, but the relation is basically flat at the low-mass end.
Reproducing and physically modeling the scatter in the mass--size relation will be the subject of further investigation.

The study of the halo--galaxy connection would be incomplete if the relations between the properties of haloes/galaxies and their spatial distribution in the LSS were not taken into account. In the last part of this paper we show that the clustering of our predicted central galaxies, measured in terms of their power spectra, reproduces that of the true sample with a high level of accuracy: 
$0.05 - 7.6 \%$, $\chi^2 = 0.18 - 8.88$ for stellar mass; 
$2.0 - 5.1 \%$, $\chi^2 = 1.9 - 7.6$ for sSFR; 
$0.12 - 11.7 \%$,  $\chi^2 = 0.06 - 13.99$ for radius; $4.4 - 7.3 \%$,  $\chi^2 = 6.7 - 15.9$ for colour. 
Importantly, this good agreement is obtained for multiple subsets defined in terms of the aforementioned galaxy properties. Despite this good performance, some subsets 
display a few percent bias (difference) in the amplitude of the spectrum. As an example, the high-mass subpopulation, 
$\log_{10}$ (M$_\ast{}[h^{-1} {\rm{M_\odot}}]) > 10.6$, and the high-sSFR subpopulation, $\log_{10} ( {\rm sSFR}[{\rm yr^{-1}} h] ) >- 11$, are predicted to have slightly smaller bias than the real TNG300 galaxies.
On the other hand, the subpopulation with large radius, $\log_{10} ( R_{1/2}^{(*)}[h^{-1} \, {\rm kpc}] ) > 0.9 $, and that with bluer colours, $(g - i) < 1$, show some scatter, but no significant bias, especially on small scales ($k\gtrsim 0.1 h$ Mpc$^{-1}$), which may point towards either hidden variables that correlate with these properties, or a larger role of stochasticity. In terms of the clustering properties of IllustrisTNG galaxies, one interesting aspect that merits further investigation, particularly if larger boxes become available, is whether clustering can be reproduced {\it{at fixed halo mass}}.

The results presented in this paper come with another important realization. 
Even though we are equipped with a powerful ML machinery and the SMOGN augmentation technique, the accuracy in the predictions for galaxy radius, sSFR, and colour is still not comparable to that of stellar mass. In Fig. \ref{fig:including_gal_props}, we show the effect of considering galaxy properties as input features (using only NN), i.e., we use all halo and galaxy properties (except for the one under analysis) to predict stellar mass, sSFR, radius, and colour. 
This exercise is reassuring in terms of the robustness of our methodology, since our predictions, in most cases, improve significantly (both qualitatively and quantitatively). 
That is the case for sSFR and colour (stellar mass was already very well reproduced), which is of course expected due to their correlation. 
Fig. \ref{fig:including_gal_props}, however, illustrates some of the challenges: even with the aid of galaxy properties, there is a substantial level of scatter which cannot be overcome with this set of properties alone. For sSFR, the scatter becomes very small for the bulk of the distribution, but there are still problems at the low-sSFR range -- a testament to the challenge of dealing with extremely low-SFR objects in TNG300. 
For colour, conversely, the scatter is larger in the blue cloud than in the red sequence. Finally, particularly striking is the effect on galaxy size, where little or no improvement is observed in the performance scores nor in the visual appearance of the scatter distribution. 
Further investigation will be devoted to improving the ML predictions for properties such as sSFR, colour, or size.

An interesting point of debate is whether the predictions for these properties can be improved upon significantly by including any additional halo or environmental property (or even information on assembly history such as the number of major mergers).
Equivalently, one can ask to which extent the problem is dominated by the known stochasticity of the galaxy formation process. 
In this sense, we have tweaked our definition of environmental overdensity, by varying the threshold scale.
We have also tested the addition of the anisotropy parameter ($\alpha$, e.g., \citealt{Paranjape2018, RamakrishnanParanjape2019} and thereafter) which, rather than describing the density itself, can be used as a descriptor of the the anisotropy of the mass distribution around haloes (i.e., whether haloes are located in nodes or filaments in the cosmic web). 
The $\alpha$ parameter has been recently proposed as a candidate to explain the halo assembly bias signal, i.e., the dependence of halo clustering on halo properties beyond halo mass.
Finally, we have also explored the use of the parametrized shape of the halo mass accretion history (MAH) as an alternative proxy for halo age -- the $\beta$ parameter, which corresponds to the slope of the MAH, see \citealt{MonteroDorta2021B}. 
This parameter has been shown to provide a more stable link to galaxy properties, at fixed halo mass, than other widely used halo age proxies such as formation time \citep{MonteroDorta2021B}. 
None of these potential refinements, however, provided significant improvements in our performance scores with respect to the basic set of halo properties, so we have opted to stick to our fiducial configuration for simplicity. 
The inclusion of more sophisticated secondary halo properties will be the subject of further investigation.

In the same context, the fact that we are able to predict the TNG300 clustering with precision, even when the sample is split in multiple ways, serves as a motivation to explore in more detail the related effect of galaxy assembly bias, i.e., the dependence of the properties and clustering of galaxies on halo properties beyond halo mass\footnote{Note that alternative definitions are common in the literature, see, e.g., \citealt{Artale2018,Zehavi2018,Bose2019,Obuljen2019}.} (see, e.g., \citealt{ShethTormen2004,Gao2005, Dalal2008,Borzyszkowski2017,Salcedo2018,SatoPolito2019,MonteroDorta2020, Tucci2021, MonteroDorta2021B}). 
On the cosmology front, as it has been already shown in the literature, ML can be used to generate high-fidelity mocks for upcoming surveys by applying the trained machine to N-body numerical simulations (e.g., \citealt{Xu2013, Zhang2019, Yip2019, renan2020, Kasmanoff2020}). 
The process can also be reverted, so that a ML method can be used to infer halo properties from galaxy observations (e.g., \citealt{Ntampaka2019, Marttens2021}). 
Finally, we will explore the application of ML in the context of {\it{multi-tracer}} clustering techniques, which are designed to reduce the uncertainties in clustering measurements by combining information from multiple galaxy populations (e.g., \citealt{Abramo2013, MTPK,MonteroDorta2020_MTPK}).   

\section*{Acknowledgements}

NSMS, NVNR, LRA and BT would like to thank São Paulo Research Foundation (FAPESP), Brazilian National Council for Scientific and Technological Development (CNPq) and Coordination for the Improvement of Higher Education Personnel (CAPES) for financial support. NSMS acknowledges financial support from FAPESP, grant \href{https://bv.fapesp.br/pt/bolsas/187647/matrizes-de-covariancia-cosmologicas-e-metodos-de-machine-learning/}{2019/13108-0}. ADMD thanks Fondecyt for financial support through the Fondecyt Regular 2021 grant 1210612. MCA acknowledges financial support from the Austrian National Science Foundation through FWF stand-alone grant P31154-N27, and MSCA Seal of Excellence @UNIPD 2020 programme under the ACROGAL project.

\section*{Data Availability}

The ML models developed in this work, as well their usage on TNG300 data and the SMOGN augmentation codes are available at \href{https://github.com/natalidesanti/mimicking_hg_connection_using_ml}{https://github.com/natalidesanti/mimicking\_hg\_connection\_using\_ml}.



\bibliographystyle{mnras}
\bibliography{example} 

\begin{thebibliography}{}
\makeatletter
\relax
\def\mn@urlcharsother{\let\do\@makeother \do\$\do\&\do\#\do\^\do\_\do\%\do\~}
\def\mn@doi{\begingroup\mn@urlcharsother \@ifnextchar [ {\mn@doi@}
  {\mn@doi@[]}}
\def\mn@doi@[#1]#2{\def\@tempa{#1}\ifx\@tempa\@empty \href
  {http://dx.doi.org/#2} {doi:#2}\else \href {http://dx.doi.org/#2} {#1}\fi
  \endgroup}
\def\mn@eprint#1#2{\mn@eprint@#1:#2::\@nil}
\def\mn@eprint@arXiv#1{\href {http://arxiv.org/abs/#1} {{\tt arXiv:#1}}}
\def\mn@eprint@dblp#1{\href {http://dblp.uni-trier.de/rec/bibtex/#1.xml}
  {dblp:#1}}
\def\mn@eprint@#1:#2:#3:#4\@nil{\def\@tempa {#1}\def\@tempb {#2}\def\@tempc
  {#3}\ifx \@tempc \@empty \let \@tempc \@tempb \let \@tempb \@tempa \fi \ifx
  \@tempb \@empty \def\@tempb {arXiv}\fi \@ifundefined
  {mn@eprint@\@tempb}{\@tempb:\@tempc}{\expandafter \expandafter \csname
  mn@eprint@\@tempb\endcsname \expandafter{\@tempc}}}

\bibitem[\protect\citeauthoryear{{Abramo} \& {Leonard}}{{Abramo} \&
  {Leonard}}{2013}]{Abramo2013}
{Abramo} L.~R.,  {Leonard} K.~E.,  2013, \mn@doi [\mnras]
  {10.1093/mnras/stt465}, \href
  {https://ui.adsabs.harvard.edu/abs/2013MNRAS.432..318A} {432, 318}

\bibitem[\protect\citeauthoryear{{Abramo}, {Secco}  \& {Loureiro}}{{Abramo}
  et~al.}{2016}]{MTPK}
{Abramo} L.~R.,  {Secco} L.~F.,   {Loureiro} A.,  2016, \mn@doi [\mnras]
  {10.1093/mnras/stv2588}, \href
  {https://ui.adsabs.harvard.edu/abs/2016MNRAS.455.3871A} {455, 3871}

\bibitem[\protect\citeauthoryear{Agarwal, Davé  \& Bassett}{Agarwal
  et~al.}{2018}]{Agarwal2018}
Agarwal S.,  Davé R.,   Bassett B.~A.,  2018, Monthly Notices of the Royal
  Astronomical Society, 478, 3410–3422

\bibitem[\protect\citeauthoryear{{Alves de Oliveira}, {Li},
  {Villaescusa-Navarro}, {Ho}  \& {Spergel}}{{Alves de Oliveira}
  et~al.}{2020}]{renan2020}
{Alves de Oliveira} R.,  {Li} Y.,  {Villaescusa-Navarro} F.,  {Ho} S.,
  {Spergel} D.~N.,  2020, arXiv e-prints, \href
  {https://ui.adsabs.harvard.edu/abs/2020arXiv201200240A} {p. arXiv:2012.00240}

\bibitem[\protect\citeauthoryear{{Artale}, {Zehavi}, {Contreras}  \&
  {Norberg}}{{Artale} et~al.}{2018}]{Artale2018}
{Artale} M.~C.,  {Zehavi} I.,  {Contreras} S.,   {Norberg} P.,  2018, \mn@doi
  [\mnras] {10.1093/mnras/sty2110}, \href
  {https://ui.adsabs.harvard.edu/abs/2018MNRAS.480.3978A} {480, 3978}

\bibitem[\protect\citeauthoryear{Awad \& Khanna}{Awad \&
  Khanna}{2015}]{Awad2015}
Awad M.,  Khanna R.,  2015, Efficient Learning Machines: Theories, Concepts,
  and Applications for Engineers and System Designers.
Apress, \url {https://doi.org/10.1007/978-1-4302-5990-9_1}

\bibitem[\protect\citeauthoryear{{Becker}}{{Becker}}{2015}]{becker2015_eam}
{Becker} M.~R.,  2015, arXiv:1507.03605, \href
  {https://ui.adsabs.harvard.edu/abs/2015arXiv150703605B} {p. arXiv:1507.03605}

\bibitem[\protect\citeauthoryear{{Behroozi}, {Conroy}  \&
  {Wechsler}}{{Behroozi} et~al.}{2010}]{Behroozi2010}
{Behroozi} P.~S.,  {Conroy} C.,   {Wechsler} R.~H.,  2010, \mn@doi [\apj]
  {10.1088/0004-637X/717/1/379}, \href
  {https://ui.adsabs.harvard.edu/abs/2010ApJ...717..379B} {717, 379}

\bibitem[\protect\citeauthoryear{{Behroozi}, {Wechsler}, {Hearin}  \&
  {Conroy}}{{Behroozi} et~al.}{2019}]{behroozi2019_um}
{Behroozi} P.,  {Wechsler} R.~H.,  {Hearin} A.~P.,   {Conroy} C.,  2019,
  \mn@doi [\mnras] {10.1093/mnras/stz1182}, \href
  {https://ui.adsabs.harvard.edu/abs/2019MNRAS.488.3143B} {488, 3143}

\bibitem[\protect\citeauthoryear{{Beltz-Mohrmann}, {Berlind}  \&
  {Szewciw}}{{Beltz-Mohrmann} et~al.}{2020}]{Beltz-Mohrmann2020}
{Beltz-Mohrmann} G.~D.,  {Berlind} A.~A.,   {Szewciw} A.~O.,  2020, \mn@doi
  [\mnras] {10.1093/mnras/stz3442}, \href
  {https://ui.adsabs.harvard.edu/abs/2020MNRAS.491.5771B} {491, 5771}

\bibitem[\protect\citeauthoryear{Berlind \& Weinberg}{Berlind \&
  Weinberg}{2002}]{Berlind2002}
Berlind A.~A.,  Weinberg D.~H.,  2002, Astrophysics Journal, 575, 587

\bibitem[\protect\citeauthoryear{Bishop}{Bishop}{1995}]{bishop1995neural}
Bishop C.,  1995, Neural Networks for Pattern Recognition.
Advanced Texts in Econometrics, Clarendon Press, \url
  {https://books.google.com.br/books?id=T0S0BgAAQBAJ}

\bibitem[\protect\citeauthoryear{Borzyszkowski, Porciani, Romano-Díaz  \&
  Garaldi}{Borzyszkowski et~al.}{2017}]{Borzyszkowski2017}
Borzyszkowski M.,  Porciani C.,  Romano-Díaz E.,   Garaldi E.,  2017, \mn@doi
  [MNRAS] {10.1093/mnras/stx873}, 469, 594–611

\bibitem[\protect\citeauthoryear{{Bose}, {Eisenstein}, {Hernquist},
  {Pillepich}, {Nelson}, {Marinacci}, {Springel}  \& {Vogelsberger}}{{Bose}
  et~al.}{2019}]{Bose2019}
{Bose} S.,  {Eisenstein} D.~J.,  {Hernquist} L.,  {Pillepich} A.,  {Nelson} D.,
   {Marinacci} F.,  {Springel} V.,   {Vogelsberger} M.,  2019, \mn@doi [\mnras]
  {10.1093/mnras/stz2546}, \href
  {https://ui.adsabs.harvard.edu/abs/2019MNRAS.tmp.2192B} {p.~2192}

\bibitem[\protect\citeauthoryear{Branco, Torgo  \& Ribeiro}{Branco
  et~al.}{2017}]{pmlr-v74-branco17a}
Branco P.,  Torgo L.,   Ribeiro R.~P.,  2017, in Torgo L.,  Krawczyk B.,
  Branco P.,   Moniz N.,  eds,  Proceedings of Machine Learning Research Vol.
  74, Proceedings of the First International Workshop on Learning with
  Imbalanced Domains: Theory and Applications. PMLR, ECML-PKDD, Skopje,
  Macedonia, pp 36--50, \url {http://proceedings.mlr.press/v74/branco17a.html}

\bibitem[\protect\citeauthoryear{Breiman}{Breiman}{1996}]{Breiman1996}
Breiman L.,  1996, \mn@doi [Machine Learning] {10.1007/BF00117832}, 24

\bibitem[\protect\citeauthoryear{Breiman}{Breiman}{2001}]{breiman2001}
Breiman L.,  2001, \mn@doi [Machine Learning]
  {https://doi.org/10.1023/A:1010933404324}, 45, 3871

\bibitem[\protect\citeauthoryear{{Bullock}, {Dekel}, {Kolatt}, {Kravtsov},
  {Klypin}, {Porciani}  \& {Primack}}{{Bullock} et~al.}{2001}]{Bullock2001_2}
{Bullock} J.~S.,  {Dekel} A.,  {Kolatt} T.~S.,  {Kravtsov} A.~V.,  {Klypin}
  A.~A.,  {Porciani} C.,   {Primack} J.~R.,  2001, \mn@doi [\apj]
  {10.1086/321477}, \href
  {https://ui.adsabs.harvard.edu/abs/2001ApJ...555..240B} {555, 240}

\bibitem[\protect\citeauthoryear{{Buser}}{{Buser}}{1978}]{Buser1978}
{Buser} R.,  1978, \aap, \href
  {https://ui.adsabs.harvard.edu/abs/1978A&A....62..411B} {62, 411}

\bibitem[\protect\citeauthoryear{Calderon \& Berlind}{Calderon \&
  Berlind}{2019}]{Calderon2019}
Calderon V.~F.,  Berlind A.~A.,  2019, Monthly Notices of the Royal
  Astronomical Society, 490, 2367–2379

\bibitem[\protect\citeauthoryear{{Carvajal}, {Matute}, {Afonso}, {Amarantidis},
  {Barbosa}, {Cunha}  \& {Humphrey}}{{Carvajal} et~al.}{2021}]{Carvajal2021}
{Carvajal} R.,  {Matute} I.,  {Afonso} J.,  {Amarantidis} S.,  {Barbosa} D.,
  {Cunha} P.,   {Humphrey} A.,  2021, \mn@doi [Galaxies]
  {10.3390/galaxies9040086}, \href
  {https://ui.adsabs.harvard.edu/abs/2021Galax...9...86C} {9, 86}

\bibitem[\protect\citeauthoryear{Chawla, Bowyer, Hall  \& Kegelmeyer}{Chawla
  et~al.}{2002}]{Chawla_2002}
Chawla N.~V.,  Bowyer K.~W.,  Hall L.~O.,   Kegelmeyer W.~P.,  2002, \mn@doi
  [Journal of Artificial Intelligence Research] {10.1613/jair.953}, 16,
  321–357

\bibitem[\protect\citeauthoryear{Chollet}{Chollet}{2015}]{chollet2015}
Chollet F.,  2015, keras, \url{https://github.com/fchollet/keras}

\bibitem[\protect\citeauthoryear{Chollet}{Chollet}{2017}]{chollet2017}
Chollet F.,  2017, Deep Learning with Python.
Manning Publications Company, \url
  {https://books.google.com.br/books?id=Yo3CAQAACAAJ}

\bibitem[\protect\citeauthoryear{{Conroy}, {Wechsler}  \& {Kravtsov}}{{Conroy}
  et~al.}{2006}]{Conroy2006}
{Conroy} C.,  {Wechsler} R.~H.,   {Kravtsov} A.~V.,  2006, \mn@doi [\apj]
  {10.1086/503602}, \href
  {https://ui.adsabs.harvard.edu/abs/2006ApJ...647..201C} {647, 201}

\bibitem[\protect\citeauthoryear{{Contreras}, {Angulo}  \&
  {Zennaro}}{{Contreras} et~al.}{2020a}]{Contreras2020}
{Contreras} S.,  {Angulo} R.,   {Zennaro} M.,  2020a, arXiv e-prints, \href
  {https://ui.adsabs.harvard.edu/abs/2020arXiv200503672C} {p. arXiv:2005.03672}

\bibitem[\protect\citeauthoryear{{Contreras}, {Angulo}  \&
  {Zennaro}}{{Contreras} et~al.}{2020b}]{Contreras2021}
{Contreras} S.,  {Angulo} R.,   {Zennaro} M.,  2020b, arXiv e-prints, \href
  {https://ui.adsabs.harvard.edu/abs/2020arXiv201206596C} {p. arXiv:2012.06596}

\bibitem[\protect\citeauthoryear{{Dalal}, {White}, {Bond}  \&
  {Shirokov}}{{Dalal} et~al.}{2008}]{Dalal2008}
{Dalal} N.,  {White} M.,  {Bond} J.~R.,   {Shirokov} A.,  2008, \mn@doi [\apj]
  {10.1086/591512}, \href
  {https://ui.adsabs.harvard.edu/abs/2008ApJ...687...12D} {687, 12}

\bibitem[\protect\citeauthoryear{Dattilo et~al.,}{Dattilo
  et~al.}{2019}]{Dattilo2019}
Dattilo A.,  et~al., 2019, \mn@doi [The Astronomical Journal]
  {10.3847/1538-3881/ab0e12}, 157, 169

\bibitem[\protect\citeauthoryear{{Davis}, {Efstathiou}, {Frenk}  \&
  {White}}{{Davis} et~al.}{1985}]{Davis1985}
{Davis} M.,  {Efstathiou} G.,  {Frenk} C.~S.,   {White} S.~D.~M.,  1985,
  \mn@doi [\apj] {10.1086/163168}, \href
  {https://ui.adsabs.harvard.edu/abs/1985ApJ...292..371D} {292, 371}

\bibitem[\protect\citeauthoryear{{Delgado}, {Wadekar}, {Hadzhiyska}, {Bose},
  {Hernquist}  \& {Ho}}{{Delgado} et~al.}{2021}]{Delgado2021}
{Delgado} A.~M.,  {Wadekar} D.,  {Hadzhiyska} B.,  {Bose} S.,  {Hernquist} L.,
   {Ho} S.,  2021, arXiv e-prints, \href
  {https://ui.adsabs.harvard.edu/abs/2021arXiv211102422D} {p. arXiv:2111.02422}

\bibitem[\protect\citeauthoryear{{Dolag}, {Borgani}, {Murante}  \&
  {Springel}}{{Dolag} et~al.}{2009}]{Dolag2009}
{Dolag} K.,  {Borgani} S.,  {Murante} G.,   {Springel} V.,  2009, \mn@doi
  [\mnras] {10.1111/j.1365-2966.2009.15034.x}, \href
  {https://ui.adsabs.harvard.edu/abs/2009MNRAS.399..497D} {399, 497}

\bibitem[\protect\citeauthoryear{D’Isanto \& Polsterer}{D’Isanto \&
  Polsterer}{2018}]{D_Isanto_2018}
D’Isanto A.,  Polsterer K.~L.,  2018, \mn@doi [Astronomy & Astrophysics]
  {10.1051/0004-6361/201731326}, 609, A111

\bibitem[\protect\citeauthoryear{{Fasano} \& {Franceschini}}{{Fasano} \&
  {Franceschini}}{1987}]{2DKS-Fasano1987}
{Fasano} G.,  {Franceschini} A.,  1987, \mn@doi [\mnras]
  {10.1093/mnras/225.1.155}, \href
  {https://ui.adsabs.harvard.edu/abs/1987MNRAS.225..155F} {225, 155}

\bibitem[\protect\citeauthoryear{{Favole} et~al.,}{{Favole}
  et~al.}{2016}]{Favole2016}
{Favole} G.,  et~al., 2016, \mn@doi [\mnras] {10.1093/mnras/stw1483}, \href
  {http://adsabs.harvard.edu/abs/2016MNRAS.461.3421F} {461, 3421}

\bibitem[\protect\citeauthoryear{{Favole}, {Montero-Dorta}, {Artale},
  {Contreras}, {Zehavi}  \& {Xu}}{{Favole} et~al.}{2021}]{Favole2021}
{Favole} G.,  {Montero-Dorta} A.~D.,  {Artale} M.~C.,  {Contreras} S.,
  {Zehavi} I.,   {Xu} X.,  2021, arXiv e-prints, \href
  {https://ui.adsabs.harvard.edu/abs/2021arXiv210110733F} {p. arXiv:2101.10733}

\bibitem[\protect\citeauthoryear{{Gao}, {Springel}  \& {White}}{{Gao}
  et~al.}{2005}]{Gao2005}
{Gao} L.,  {Springel} V.,   {White} S.~D.~M.,  2005, \mn@doi [\mnras]
  {10.1111/j.1745-3933.2005.00084.x}, \href
  {http://adsabs.harvard.edu/abs/2005MNRAS.363L..66G} {363, L66}

\bibitem[\protect\citeauthoryear{{Genel} et~al.,}{{Genel}
  et~al.}{2014}]{Genel2014}
{Genel} S.,  et~al., 2014, \mn@doi [\mnras] {10.1093/mnras/stu1654}, \href
  {https://ui.adsabs.harvard.edu/abs/2014MNRAS.445..175G} {445, 175}

\bibitem[\protect\citeauthoryear{Geurts, Ernst  \& Wehenkel}{Geurts
  et~al.}{2006}]{ert_paper}
Geurts P.,  Ernst D.,   Wehenkel L.,  2006, \mn@doi [Machine Learning]
  {10.1007/s10994-006-6226-1}, 63, 3

\bibitem[\protect\citeauthoryear{Golob, Sawicki, Goulding  \& Coupon}{Golob
  et~al.}{2021}]{golob2021}
Golob A.,  Sawicki M.,  Goulding A.~D.,   Coupon J.,  2021, \mn@doi [Monthly
  Notices of the Royal Astronomical Society] {10.1093/mnras/stab719}, 503,
  4136–4146

\bibitem[\protect\citeauthoryear{{Gu} et~al.,}{{Gu} et~al.}{2020}]{Gu2020}
{Gu} M.,  et~al., 2020, arXiv e-prints, \href
  {https://ui.adsabs.harvard.edu/abs/2020arXiv201004166G} {p. arXiv:2010.04166}

\bibitem[\protect\citeauthoryear{Guo, White, Angulo, Henriques, Lemson,
  Boylan-Kolchin, Thomas  \& Short}{Guo et~al.}{2013}]{Guo2013}
Guo Q.,  White S.,  Angulo R.,  Henriques B.,  Lemson G.,  Boylan-Kolchin M.,
  Thomas P.,   Short C.,  2013, Monthly Notices of the Royal Astronomical
  Society, 428, 1351

\bibitem[\protect\citeauthoryear{{Guo} et~al.,}{{Guo} et~al.}{2016}]{Guo2016}
{Guo} H.,  et~al., 2016, \mn@doi [\mnras] {10.1093/mnras/stw845}, \href
  {https://ui.adsabs.harvard.edu/abs/2016MNRAS.459.3040G} {459, 3040}

\bibitem[\protect\citeauthoryear{{Hadzhiyska}, {Bose}, {Eisenstein}  \&
  {Hernquist}}{{Hadzhiyska} et~al.}{2020a}]{Hadzhiyska2020B}
{Hadzhiyska} B.,  {Bose} S.,  {Eisenstein} D.,   {Hernquist} L.,  2020a, arXiv
  e-prints, \href {https://ui.adsabs.harvard.edu/abs/2020arXiv200804913H} {p.
  arXiv:2008.04913}

\bibitem[\protect\citeauthoryear{{Hadzhiyska}, {Bose}, {Eisenstein},
  {Hernquist}  \& {Spergel}}{{Hadzhiyska} et~al.}{2020b}]{Hadzhiyska2020}
{Hadzhiyska} B.,  {Bose} S.,  {Eisenstein} D.,  {Hernquist} L.,   {Spergel}
  D.~N.,  2020b, \mn@doi [\mnras] {10.1093/mnras/staa623}, \href
  {https://ui.adsabs.harvard.edu/abs/2020MNRAS.493.5506H} {493, 5506}

\bibitem[\protect\citeauthoryear{{Hadzhiyska}, {Bose}, {Eisenstein}  \&
  {Hernquist}}{{Hadzhiyska} et~al.}{2021}]{Hadzhiyska2021}
{Hadzhiyska} B.,  {Bose} S.,  {Eisenstein} D.,   {Hernquist} L.,  2021, \mn@doi
  [\mnras] {10.1093/mnras/staa3776}, \href
  {https://ui.adsabs.harvard.edu/abs/2021MNRAS.501.1603H} {501, 1603}

\bibitem[\protect\citeauthoryear{{Hand}, {Feng}, {Beutler}, {Li}, {Modi},
  {Seljak}  \& {Slepian}}{{Hand} et~al.}{2018}]{nbodykit2018}
{Hand} N.,  {Feng} Y.,  {Beutler} F.,  {Li} Y.,  {Modi} C.,  {Seljak} U.,
  {Slepian} Z.,  2018, \mn@doi [\aj] {10.3847/1538-3881/aadae0}, \href
  {https://ui.adsabs.harvard.edu/abs/2018AJ....156..160H} {156, 160}

\bibitem[\protect\citeauthoryear{Ivezi{\'c}, Connolly, VanderPlas  \&
  Gray}{Ivezi{\'c} et~al.}{2014}]{ivezic2014statistics}
Ivezi{\'c} {\v{Z}}.,  Connolly A.,  VanderPlas J.,   Gray A.,  2014,
  Statistics, Data Mining, and Machine Learning in Astronomy: A Practical
  Python Guide for the Analysis of Survey Data.
Princeton Series in Modern Observational Astronomy, Princeton University Press,
  \url {https://books.google.com.br/books?id=h2eYDwAAQBAJ}

\bibitem[\protect\citeauthoryear{{Jo} \& {Kim}}{{Jo} \& {Kim}}{2019}]{Jo2019}
{Jo} Y.,  {Kim} J.-h.,  2019, \mn@doi [\mnras] {10.1093/mnras/stz2304}, \href
  {https://ui.adsabs.harvard.edu/abs/2019MNRAS.489.3565J} {489, 3565}

\bibitem[\protect\citeauthoryear{{Kamdar}, {Turk}  \& {Brunner}}{{Kamdar}
  et~al.}{2016}]{Kamdar2016}
{Kamdar} H.~M.,  {Turk} M.~J.,   {Brunner} R.~J.,  2016, \mn@doi [\mnras]
  {10.1093/mnras/stv2981}, \href
  {https://ui.adsabs.harvard.edu/abs/2016MNRAS.457.1162K} {457, 1162}

\bibitem[\protect\citeauthoryear{{Kasmanoff}, {Villaescusa-Navarro}, {Tinker}
  \& {Ho}}{{Kasmanoff} et~al.}{2020}]{Kasmanoff2020}
{Kasmanoff} N.,  {Villaescusa-Navarro} F.,  {Tinker} J.,   {Ho} S.,  2020,
  arXiv e-prints, \href {https://ui.adsabs.harvard.edu/abs/2020arXiv201200186K}
  {p. arXiv:2012.00186}

\bibitem[\protect\citeauthoryear{Ke, Meng, Finley, Wang, Chen, Ma, Ye  \&
  Liu}{Ke et~al.}{2017}]{lgbm_paper}
Ke G.,  Meng Q.,  Finley T.,  Wang T.,  Chen W.,  Ma W.,  Ye Q.,   Liu T.-Y.,
  2017, in Guyon I.,  Luxburg U.~V.,  Bengio S.,  Wallach H.,  Fergus R.,
  Vishwanathan S.,   Garnett R.,  eds, ~ Vol. 30, {Advances in Neural
  Information Processing Systems}. Curran Associates, Inc., \url
  {https://proceedings.neurips.cc/paper/2017/file/6449f44a102fde848669bdd9eb6b76fa-Paper.pdf}

\bibitem[\protect\citeauthoryear{Krawczyk}{Krawczyk}{2016}]{Krawczyk2016}
Krawczyk B.,  2016, \mn@doi [Progress in Artificial Intelligence]
  {10.1007/s13748-016-0094-0}, 5

\bibitem[\protect\citeauthoryear{Kunz}{Kunz}{2019}]{Kunz2019}
Kunz N.,  2019, SMOGN, \url{https://github.com/nickkunz/smogn}

\bibitem[\protect\citeauthoryear{{Li} et~al.,}{{Li} et~al.}{2021}]{Li2021}
{Li} N.,  et~al., 2021, \mn@doi [\apj] {10.3847/1538-4357/ac0973}, \href
  {https://ui.adsabs.harvard.edu/abs/2021ApJ...917...72L} {917, 72}

\bibitem[\protect\citeauthoryear{{Lovell}, {Wilkins}, {Thomas}, {Schaller},
  {Baugh}, {Fabbian}  \& {Bah{\'e}}}{{Lovell} et~al.}{2022}]{Lovell2021}
{Lovell} C.~C.,  {Wilkins} S.~M.,  {Thomas} P.~A.,  {Schaller} M.,  {Baugh}
  C.~M.,  {Fabbian} G.,   {Bah{\'e}} Y.,  2022, \mn@doi [\mnras]
  {10.1093/mnras/stab3221}, \href
  {https://ui.adsabs.harvard.edu/abs/2022MNRAS.509.5046L} {509, 5046}

\bibitem[\protect\citeauthoryear{Lu, Ren, Li, Zhao, Xu  \& Ren}{Lu
  et~al.}{2021}]{Lu2021}
Lu J.,  Ren K.,  Li X.,  Zhao Y.,  Xu Z.,   Ren X.,  2021, \mn@doi
  [GeoInformatica] {10.1007/s10707-020-00426-7}, pp 1--32

\bibitem[\protect\citeauthoryear{{Lucie-Smith}, {Peiris}  \&
  {Pontzen}}{{Lucie-Smith} et~al.}{2019}]{Lucie2019}
{Lucie-Smith} L.,  {Peiris} H.~V.,   {Pontzen} A.,  2019, \mn@doi [\mnras]
  {10.1093/mnras/stz2599}, \href
  {https://ui.adsabs.harvard.edu/abs/2019MNRAS.490..331L} {490, 331}

\bibitem[\protect\citeauthoryear{Man, Peng, Shi, Kon, Zhang, Dou  \& Guo}{Man
  et~al.}{2019}]{Man2019}
Man Z.-Y.,  Peng Y.-J.,  Shi J.-J.,  Kon X.,  Zhang C.-P.,  Dou J.,   Guo
  K.-X.,  2019, The Astrophysical Journal, 881

\bibitem[\protect\citeauthoryear{{Marinacci} et~al.,}{{Marinacci}
  et~al.}{2018}]{Marinacci2018}
{Marinacci} F.,  et~al., 2018, \mn@doi [\mnras] {10.1093/mnras/sty2206}, \href
  {https://ui.adsabs.harvard.edu/abs/2018MNRAS.480.5113M} {480, 5113}

\bibitem[\protect\citeauthoryear{{McGibbon} \& {Khochfar}}{{McGibbon} \&
  {Khochfar}}{2021}]{McGibbon2021}
{McGibbon} R.,  {Khochfar} S.,  2021, arXiv e-prints, \href
  {https://ui.adsabs.harvard.edu/abs/2021arXiv211208424M} {p. arXiv:2112.08424}

\bibitem[\protect\citeauthoryear{{Montero-Dorta}, {Abramo}, {Granett}, {de la
  Torre}  \& {Guzzo}}{{Montero-Dorta} et~al.}{2020a}]{MonteroDorta2020_MTPK}
{Montero-Dorta} A.~D.,  {Abramo} L.~R.,  {Granett} B.~R.,  {de la Torre} S.,
  {Guzzo} L.,  2020a, \mn@doi [\mnras] {10.1093/mnras/staa405}, \href
  {https://ui.adsabs.harvard.edu/abs/2020MNRAS.493.5257M} {493, 5257}

\bibitem[\protect\citeauthoryear{{Montero-Dorta} et~al.,}{{Montero-Dorta}
  et~al.}{2020b}]{MonteroDorta2020}
{Montero-Dorta} A.~D.,  et~al., 2020b, \mn@doi [\mnras]
  {10.1093/mnras/staa1624}, \href
  {https://ui.adsabs.harvard.edu/abs/2020MNRAS.496.1182M} {496, 1182}

\bibitem[\protect\citeauthoryear{{Montero-Dorta}, {Artale}, {Abramo}  \&
  {Tucci}}{{Montero-Dorta} et~al.}{2021a}]{MonteroDorta2021A}
{Montero-Dorta} A.~D.,  {Artale} M.~C.,  {Abramo} L.~R.,   {Tucci} B.,  2021a,
  \mn@doi [\mnras] {10.1093/mnras/stab1026}, \href
  {https://ui.adsabs.harvard.edu/abs/2021MNRAS.504.4568M} {504, 4568}

\bibitem[\protect\citeauthoryear{{Montero-Dorta}, {Chaves-Montero}, {Artale}
  \& {Favole}}{{Montero-Dorta} et~al.}{2021b}]{MonteroDorta2021B}
{Montero-Dorta} A.~D.,  {Chaves-Montero} J.,  {Artale} M.~C.,   {Favole} G.,
  2021b, \mn@doi [\mnras] {10.1093/mnras/stab2556}, \href
  {https://ui.adsabs.harvard.edu/abs/2021MNRAS.508..940M} {508, 940}

\bibitem[\protect\citeauthoryear{{Moster}, {Naab}  \& {White}}{{Moster}
  et~al.}{2018}]{moster2018_emerge}
{Moster} B.~P.,  {Naab} T.,   {White} S. D.~M.,  2018, \mn@doi [\mnras]
  {10.1093/mnras/sty655}, \href
  {https://ui.adsabs.harvard.edu/abs/2018MNRAS.477.1822M} {477, 1822}

\bibitem[\protect\citeauthoryear{Mustafa, Bard, Bhimji, Lukić, Al-Rfou  \&
  Kratochvil}{Mustafa et~al.}{2019}]{Mustafa2019}
Mustafa M.,  Bard D.,  Bhimji W.,  Lukić Z.,  Al-Rfou R.,   Kratochvil J.~M.,
  2019, \mn@doi [Computational Astrophysics and Cosmology]
  {10.1186/s40668-019-0029-9}, 6

\bibitem[\protect\citeauthoryear{{Naab} \& {Ostriker}}{{Naab} \&
  {Ostriker}}{2017}]{Naab2017}
{Naab} T.,  {Ostriker} J.~P.,  2017, \mn@doi [\araa]
  {10.1146/annurev-astro-081913-040019}, \href
  {https://ui.adsabs.harvard.edu/abs/2017ARA&A..55...59N} {55, 59}

\bibitem[\protect\citeauthoryear{{Naiman} et~al.,}{{Naiman}
  et~al.}{2018}]{Naiman2018}
{Naiman} J.~P.,  et~al., 2018, \mn@doi [\mnras] {10.1093/mnras/sty618}, \href
  {https://ui.adsabs.harvard.edu/abs/2018MNRAS.477.1206N} {477, 1206}

\bibitem[\protect\citeauthoryear{{Navarro}, {Frenk}  \& {White}}{{Navarro}
  et~al.}{1997}]{nfw1997}
{Navarro} J.~F.,  {Frenk} C.~S.,   {White} S. D.~M.,  1997, \mn@doi [\apj]
  {10.1086/304888}, \href
  {https://ui.adsabs.harvard.edu/abs/1997ApJ...490..493N} {490, 493}

\bibitem[\protect\citeauthoryear{{Nelson} et~al.,}{{Nelson}
  et~al.}{2018}]{Nelson2018_ColorBim}
{Nelson} D.,  et~al., 2018, \mn@doi [\mnras] {10.1093/mnras/stx3040}, \href
  {https://ui.adsabs.harvard.edu/abs/2018MNRAS.475..624N} {475, 624}

\bibitem[\protect\citeauthoryear{Nelson, Springel, Pillepich  \& et al.}{Nelson
  et~al.}{2019}]{Nelson2019}
Nelson D.,  Springel V.,  Pillepich A.,   et al. 2019, Computational
  Astrophysics and Cosmology, 6

\bibitem[\protect\citeauthoryear{{Ntampaka} et~al.,}{{Ntampaka}
  et~al.}{2019}]{Ntampaka2019}
{Ntampaka} M.,  et~al., 2019, \mn@doi [\apj] {10.3847/1538-4357/ab14eb}, \href
  {https://ui.adsabs.harvard.edu/abs/2019ApJ...876...82N} {876, 82}

\bibitem[\protect\citeauthoryear{{Obuljen}, {Dalal}  \& {Percival}}{{Obuljen}
  et~al.}{2019}]{Obuljen2019}
{Obuljen} A.,  {Dalal} N.,   {Percival} W.~J.,  2019, \mn@doi [\jcap]
  {10.1088/1475-7516/2019/10/020}, \href
  {https://ui.adsabs.harvard.edu/abs/2019JCAP...10..020O} {2019, 020}

\bibitem[\protect\citeauthoryear{{Paranjape}, {Hahn}  \& {Sheth}}{{Paranjape}
  et~al.}{2018}]{Paranjape2018}
{Paranjape} A.,  {Hahn} O.,   {Sheth} R.~K.,  2018, \mn@doi [\mnras]
  {10.1093/mnras/sty496}, \href
  {https://ui.adsabs.harvard.edu/abs/2018MNRAS.476.3631P} {476, 3631}

\bibitem[\protect\citeauthoryear{{Peacock}}{{Peacock}}{1983}]{2DKS-Peacock1983}
{Peacock} J.~A.,  1983, \mn@doi [\mnras] {10.1093/mnras/202.3.615}, \href
  {https://ui.adsabs.harvard.edu/abs/1983MNRAS.202..615P} {202, 615}

\bibitem[\protect\citeauthoryear{Pedregosa et~al.,}{Pedregosa
  et~al.}{2011}]{scikit-learn}
Pedregosa F.,  et~al., 2011, Journal of Machine Learning Research, 12, 2825

\bibitem[\protect\citeauthoryear{{Pillepich} et~al.,}{{Pillepich}
  et~al.}{2018a}]{Pillepich2018}
{Pillepich} A.,  et~al., 2018a, \mn@doi [\mnras] {10.1093/mnras/stx2656}, \href
  {https://ui.adsabs.harvard.edu/abs/2018MNRAS.473.4077P} {473, 4077}

\bibitem[\protect\citeauthoryear{{Pillepich} et~al.,}{{Pillepich}
  et~al.}{2018b}]{Pillepich2018b}
{Pillepich} A.,  et~al., 2018b, \mn@doi [\mnras] {10.1093/mnras/stx3112}, \href
  {https://ui.adsabs.harvard.edu/abs/2018MNRAS.475..648P} {475, 648}

\bibitem[\protect\citeauthoryear{{Planck Collaboration} et~al.,}{{Planck
  Collaboration} et~al.}{2016}]{Planck2016}
{Planck Collaboration} et~al., 2016, \mn@doi [\aap]
  {10.1051/0004-6361/201525830}, \href
  {https://ui.adsabs.harvard.edu/abs/2016A&A...594A..13P} {594, A13}

\bibitem[\protect\citeauthoryear{Ramakrishnan, Paranjape, Hahn  \&
  Sheth}{Ramakrishnan et~al.}{2019}]{RamakrishnanParanjape2019}
Ramakrishnan S.,  Paranjape A.,  Hahn O.,   Sheth R.~K.,  2019, \mn@doi
  [Monthly Notices of the Royal Astronomical Society] {10.1093/mnras/stz2344},
  489, 2977

\bibitem[\protect\citeauthoryear{{Rodriguez}, {Montero-Dorta}, {Angulo},
  {Artale}  \& {Merch{\'a}n}}{{Rodriguez} et~al.}{2021}]{Rodriguez2021}
{Rodriguez} F.,  {Montero-Dorta} A.~D.,  {Angulo} R.~E.,  {Artale} M.~C.,
  {Merch{\'a}n} M.,  2021, \mn@doi [\mnras] {10.1093/mnras/stab1571}, \href
  {https://ui.adsabs.harvard.edu/abs/2021MNRAS.505.3192R} {505, 3192}

\bibitem[\protect\citeauthoryear{{Salcedo}, {Maller}, {Berlind}, {Sinha},
  {McBride}, {Behroozi}, {Wechsler}  \& {Weinberg}}{{Salcedo}
  et~al.}{2018}]{Salcedo2018}
{Salcedo} A.~N.,  {Maller} A.~H.,  {Berlind} A.~A.,  {Sinha} M.,  {McBride}
  C.~K.,  {Behroozi} P.~S.,  {Wechsler} R.~H.,   {Weinberg} D.~H.,  2018,
  \mn@doi [\mnras] {10.1093/mnras/sty109}, \href
  {https://ui.adsabs.harvard.edu/abs/2018MNRAS.475.4411S} {475, 4411}

\bibitem[\protect\citeauthoryear{{Sato-Polito}, {Montero-Dorta}, {Abramo},
  {Prada}  \& {Klypin}}{{Sato-Polito} et~al.}{2019}]{SatoPolito2019}
{Sato-Polito} G.,  {Montero-Dorta} A.~D.,  {Abramo} L.~R.,  {Prada} F.,
  {Klypin} A.,  2019, \mn@doi [\mnras] {10.1093/mnras/stz1338}, \href
  {https://ui.adsabs.harvard.edu/abs/2019MNRAS.487.1570S} {487, 1570}

\bibitem[\protect\citeauthoryear{{Shao} et~al.,}{{Shao}
  et~al.}{2021}]{shao2021}
{Shao} H.,  et~al., 2021, arXiv e-prints, \href
  {https://ui.adsabs.harvard.edu/abs/2021arXiv210904484S} {p. arXiv:2109.04484}

\bibitem[\protect\citeauthoryear{{Sharif}, {Marijan}  \& {Liaaen}}{{Sharif}
  et~al.}{2021}]{Sharif2021}
{Sharif} A.,  {Marijan} D.,   {Liaaen} M.,  2021, arXiv e-prints, \href
  {https://ui.adsabs.harvard.edu/abs/2021arXiv211007443S} {p. arXiv:2110.07443}

\bibitem[\protect\citeauthoryear{{Sheth} \& {Tormen}}{{Sheth} \&
  {Tormen}}{2004}]{ShethTormen2004}
{Sheth} R.~K.,  {Tormen} G.,  2004, \mn@doi [\mnras]
  {10.1111/j.1365-2966.2004.07733.x}, \href
  {https://ui.adsabs.harvard.edu/abs/2004MNRAS.350.1385S} {350, 1385}

\bibitem[\protect\citeauthoryear{{Shi} et~al.,}{{Shi} et~al.}{2020}]{Shi2020}
{Shi} J.,  et~al., 2020, \mn@doi [\apj] {10.3847/1538-4357/ab8464}, \href
  {https://ui.adsabs.harvard.edu/abs/2020ApJ...893..139S} {893, 139}

\bibitem[\protect\citeauthoryear{{Somerville} \& {Dav{\'e}}}{{Somerville} \&
  {Dav{\'e}}}{2015}]{Somerville2015}
{Somerville} R.~S.,  {Dav{\'e}} R.,  2015, \mn@doi [\araa]
  {10.1146/annurev-astro-082812-140951}, \href
  {https://ui.adsabs.harvard.edu/abs/2015ARA&A..53...51S} {53, 51}

\bibitem[\protect\citeauthoryear{{Springel}}{{Springel}}{2010}]{Springel2010}
{Springel} V.,  2010, \mn@doi [\mnras] {10.1111/j.1365-2966.2009.15715.x},
  \href {https://ui.adsabs.harvard.edu/abs/2010MNRAS.401..791S} {401, 791}

\bibitem[\protect\citeauthoryear{{Springel}, {White}, {Tormen}  \&
  {Kauffmann}}{{Springel} et~al.}{2001}]{Springel2001}
{Springel} V.,  {White} S. D.~M.,  {Tormen} G.,   {Kauffmann} G.,  2001,
  \mn@doi [\mnras] {10.1046/j.1365-8711.2001.04912.x}, \href
  {https://ui.adsabs.harvard.edu/abs/2001MNRAS.328..726S} {328, 726}

\bibitem[\protect\citeauthoryear{{Springel} et~al.,}{{Springel}
  et~al.}{2018}]{Springel2018}
{Springel} V.,  et~al., 2018, \mn@doi [\mnras] {10.1093/mnras/stx3304}, \href
  {https://ui.adsabs.harvard.edu/abs/2018MNRAS.475..676S} {475, 676}

\bibitem[\protect\citeauthoryear{Taillon}{Taillon}{2018}]{2DKS}
Taillon G.,  2018, 2DKS, \url{https://github.com/Gabinou/2DKS}

\bibitem[\protect\citeauthoryear{{Trujillo-Gomez}, {Klypin}, {Primack}  \&
  {Romanowsky}}{{Trujillo-Gomez} et~al.}{2011}]{Trujillo-Gomez2011}
{Trujillo-Gomez} S.,  {Klypin} A.,  {Primack} J.,   {Romanowsky} A.~J.,  2011,
  \mn@doi [\apj] {10.1088/0004-637X/742/1/16}, \href
  {https://ui.adsabs.harvard.edu/abs/2011ApJ...742...16T} {742, 16}

\bibitem[\protect\citeauthoryear{{Tucci}, {Montero-Dorta}, {Abramo},
  {Sato-Polito}  \& {Artale}}{{Tucci} et~al.}{2021}]{Tucci2021}
{Tucci} B.,  {Montero-Dorta} A.~D.,  {Abramo} L.~R.,  {Sato-Polito} G.,
  {Artale} M.~C.,  2021, \mn@doi [\mnras] {10.1093/mnras/staa3319}, \href
  {https://ui.adsabs.harvard.edu/abs/2021MNRAS.500.2777T} {500, 2777}

\bibitem[\protect\citeauthoryear{{Villaescusa-Navarro}
  et~al.,}{{Villaescusa-Navarro} et~al.}{2021}]{VillaescusaNavarro2021}
{Villaescusa-Navarro} F.,  et~al., 2021, \mn@doi [\apj]
  {10.3847/1538-4357/abf7ba}, \href
  {https://ui.adsabs.harvard.edu/abs/2021ApJ...915...71V} {915, 71}

\bibitem[\protect\citeauthoryear{{Vogelsberger} et~al.,}{{Vogelsberger}
  et~al.}{2014a}]{Vogelsberger2014a}
{Vogelsberger} M.,  et~al., 2014a, \mn@doi [\mnras] {10.1093/mnras/stu1536},
  \href {https://ui.adsabs.harvard.edu/abs/2014MNRAS.444.1518V} {444, 1518}

\bibitem[\protect\citeauthoryear{{Vogelsberger} et~al.,}{{Vogelsberger}
  et~al.}{2014b}]{Vogelsberger2014b}
{Vogelsberger} M.,  et~al., 2014b, \mn@doi [\nat] {10.1038/nature13316}, \href
  {https://ui.adsabs.harvard.edu/abs/2014Natur.509..177V} {509, 177}

\bibitem[\protect\citeauthoryear{Wechsler \& Tinker}{Wechsler \&
  Tinker}{2018}]{Wechsler2018}
Wechsler R.~H.,  Tinker J.~L.,  2018, \mn@doi [Annual Review of Astronomy and
  Astrophysics] {10.1146/annurev-astro-081817-051756}, 56, 435–487

\bibitem[\protect\citeauthoryear{White \& Frenk}{White \&
  Frenk}{1991}]{White1991}
White S. D.~M.,  Frenk C.~S.,  1991, Astrophysical Journal, 379, 52

\bibitem[\protect\citeauthoryear{{Xu}, {Ho}, {Trac}, {Schneider}, {Poczos}  \&
  {Ntampaka}}{{Xu} et~al.}{2013}]{Xu2013}
{Xu} X.,  {Ho} S.,  {Trac} H.,  {Schneider} J.,  {Poczos} B.,   {Ntampaka} M.,
  2013, \mn@doi [\apj] {10.1088/0004-637X/772/2/147}, \href
  {https://ui.adsabs.harvard.edu/abs/2013ApJ...772..147X} {772, 147}

\bibitem[\protect\citeauthoryear{{Xu}, {Zehavi}  \& {Contreras}}{{Xu}
  et~al.}{2021}]{Xu2021}
{Xu} X.,  {Zehavi} I.,   {Contreras} S.,  2021, \mn@doi [\mnras, in press]
  {10.1093/mnras/stab100}, \href
  {https://ui.adsabs.harvard.edu/abs/2021MNRAS.tmp..129X} {}

\bibitem[\protect\citeauthoryear{{Yip} et~al.,}{{Yip} et~al.}{2019}]{Yip2019}
{Yip} J. H.~T.,  et~al., 2019, arXiv e-prints, \href
  {https://ui.adsabs.harvard.edu/abs/2019arXiv191007813Y} {p. arXiv:1910.07813}

\bibitem[\protect\citeauthoryear{{Zehavi} et~al.,}{{Zehavi}
  et~al.}{2005}]{Zehavi2005}
{Zehavi} I.,  et~al., 2005, \mn@doi [\apj] {10.1086/427495}, \href
  {http://adsabs.harvard.edu/abs/2005ApJ...621...22Z} {621, 22}

\bibitem[\protect\citeauthoryear{{Zehavi}, {Contreras}, {Padilla}, {Smith},
  {Baugh}  \& {Norberg}}{{Zehavi} et~al.}{2018}]{Zehavi2018}
{Zehavi} I.,  {Contreras} S.,  {Padilla} N.,  {Smith} N.~J.,  {Baugh} C.~M.,
  {Norberg} P.,  2018, \mn@doi [\apj] {10.3847/1538-4357/aaa54a}, \href
  {https://ui.adsabs.harvard.edu/abs/2018ApJ...853...84Z} {853, 84}

\bibitem[\protect\citeauthoryear{{Zhang}, {Wang}, {Zhang}, {Sun}, {He},
  {Contardo}, {Villaescusa-Navarro}  \& {Ho}}{{Zhang} et~al.}{2019}]{Zhang2019}
{Zhang} X.,  {Wang} Y.,  {Zhang} W.,  {Sun} Y.,  {He} S.,  {Contardo} G.,
  {Villaescusa-Navarro} F.,   {Ho} S.,  2019, arXiv e-prints, \href
  {https://ui.adsabs.harvard.edu/abs/2019arXiv190205965Z} {p. arXiv:1902.05965}

\bibitem[\protect\citeauthoryear{{von Marttens} et~al.,}{{von Marttens}
  et~al.}{2021}]{Marttens2021}
{von Marttens} R.,  et~al., 2021, arXiv e-prints, \href
  {https://ui.adsabs.harvard.edu/abs/2021arXiv211101185V} {p. arXiv:2111.01185}

\makeatother
\end{thebibliography}



\appendix

\begin{figure}
  \centering
    \includegraphics[scale=0.44]{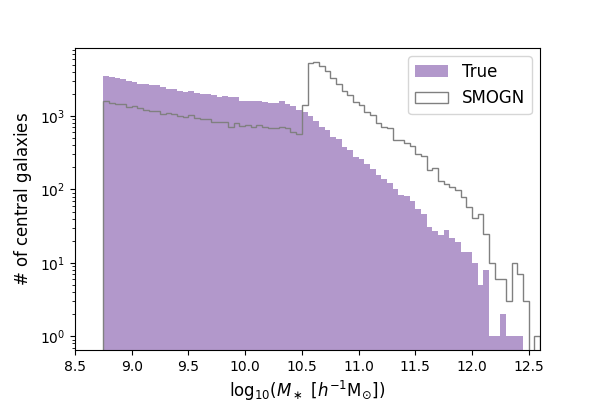}
    \includegraphics[scale=0.44]{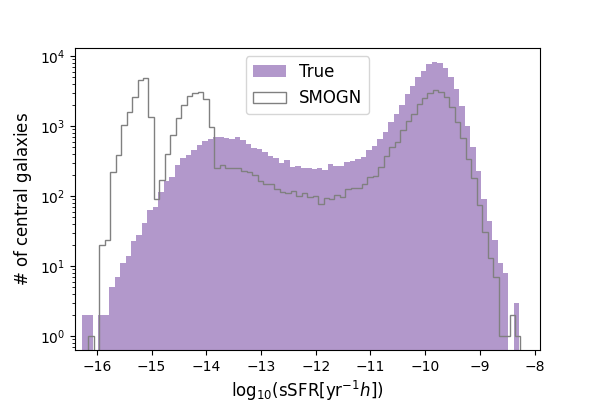}
    \includegraphics[scale=0.44]{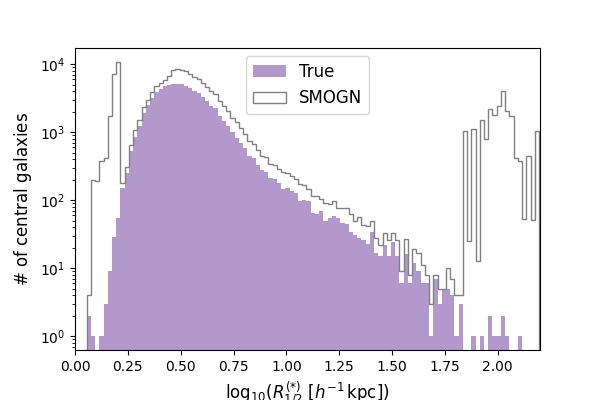}
    \includegraphics[scale=0.44]{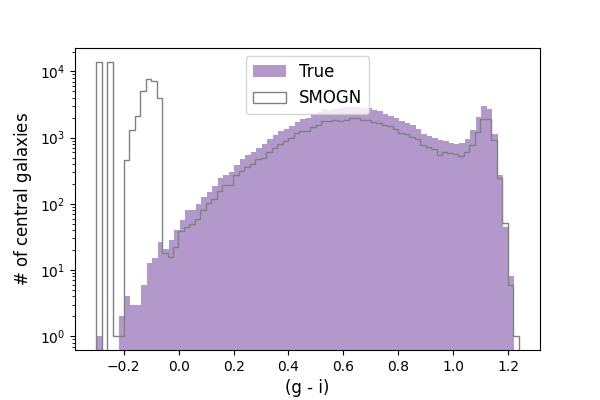}
  \caption{A comparison of the histograms for the true and augmented SMOGN distributions for stellar mass, sSFR, radius, and colour.}
    \label{fig:appendix_SMOGN}
\end{figure}

\section{The SMOGN galaxy distributions}
\label{appendix}

The SMOGN code was used in the way that, by visual inspection, we manually selected the regions of the distributions to be over- and under-sampled, according to the chosen proportion of normal and rare bins in each case and the chosen number of $k$ neighbours.
Fig. \ref{fig:appendix_SMOGN} displays the distributions of galaxy properties (stellar mass, sSFR, radius, and colour) after the SMOGN data augmentation technique has been applied on the training and validation data sets. In essence, SMOGN allows us to boost the statistics for the underrepresented populations of high stellar mass, very low sSFR, very small and very large radius, and very blue colour. Importantly, these enhanced distributions are
different from the original ones because they are designed to ``force'' the methods to learn (in the 
training stage) how to predict properties in previously underrepresented regions in parameter space.  Once the trained model is obtained, 
the test data set (which is new to the machine and not SMOGN augmented) is employed to produce the expected results: the ``complete'' (or something close to it) distribution.

The code used to produce these augmented distributions is available at the \texttt{github}'s \href{https://github.com/natalidesanti/mimicking_hg_connection_using_ml}{page}.

\section{Score comparisons}
\label{appendix_scores}

In this appendix, Table \ref{tab:score_comp}, we present the exact value obtained for MSE and PCC metrics measured in our test subset for all the different galaxy properties using the stacked models raw and SMOGN. Those results summarises the advantages of using the stacked models compared to other works (e.g., PCC $\in [0.92, 0.957]$, from \citealt{Kamdar2016, Agarwal2018, Lovell2021} for stellar mass; MSE $= 0.126$ from  \citealt{Kamdar2016} for stellar mass; PCC $ \in [0.745, 0.794]$, for SFR according to \citealt{Kamdar2016, Agarwal2018, Lovell2021}).

\begin{table}
  \caption{\label{tab:score_comp} MSE and PCC scores obtained for galaxy properties (in the test subset) for the raw and SMOGN models.}
 \begin{center}
  \begin{tabular}{c|cc|cc}
   \cline{1-5}
   \multirow{2}{*}{\textbf{Property}} & \multicolumn{2}{c|}{\textbf{Raw}} & \multicolumn{2}{|c}{\textbf{SMOGN}} \\ 
   \cline{2-5}
   & \textbf{MSE} & \textbf{PCC} & \textbf{MSE} & \textbf{PCC} \\
   \cline{1-5}
   Stellar mass & 0.017 & 0.98 & 0.018 & 0.98\\
   sSFR & 0.691 & 0.79 & 0.747 & 0.79\\
   Radius & 0.012 & 0.75 & 0.014 & 0.71\\
   Colour & 0.032 & 0.71 & 0.036 & 0.67\\
   \cline{1-5}
   \end{tabular}
  \end{center}
\end{table}

\bsp	
\label{lastpage}
\end{document}